\documentclass[toc]{PoS}

\usepackage{amsmath, array}
\usepackage{mathtools}
\usepackage{amssymb}
\usepackage{amsthm}
\usepackage{amsfonts}

\usepackage{verbatim}
\usepackage{float}
\usepackage{arydshln}

\usepackage{tikz}
\usepackage[retainorgcmds]{IEEEtrantools}

\title{Solving holographic defects}

\ShortTitle{Solving holographic defects}

\author{\speaker{Georgios Linardopoulos} \vspace{.1cm} \\
Institute of Nuclear and Particle Physics \\
N.C.S.R.\ "Demokritos" \\
153 10, Agia Paraskevi, Greece. \vspace{.1cm} \\
Department of Nuclear and Particle Physics \\
Faculty of Physics, National and Kapodistrian University of Athens \\
157 84 Athens, Greece. \vspace{.1cm} \\
E-mail: \email{glinard@inp.demokritos.gr}}

\abstract{Defect conformal field theories (dCFTs) have been attracting increased attention recently, mainly because they enable us to bridge the gap between idealistic, highly symmetric models of our world (such as the particle/string duality) and real-world systems. This talk is about the AdS/defect CFT correspondence, an exciting new proposal that joins the forces of holography, integrability, supersymmetric localization\footnote{Though the solution of defects via supersymmetric localization is only too recent to be discussed here.} \ and the conformal bootstrap program in a framework that is appropriate for the study of defects in real-world systems. After introducing dCFTs and some of their holographic realizations, we will present some recent results for the one-point functions of the integrable dCFTs that are the holographic duals of the D3-probe-D5 and the D3-probe-D7 systems of intersecting branes.}

\FullConference{Corfu Summer Institute 2019 "School and Workshops on Elementary Particle Physics and Gravity" (CORFU2019)\\
		31 August--25 September 2019\\
		Corfu, Greece}

\begin{document}
\newpage
\section[Conformal field theories with a boundary]{Conformal field theories with a boundary \label{Section:BoundaryConformalFieldTheories}}
\subsection[Conformal field theories]{Conformal field theories}
\noindent A well-known result in conformal field theory (CFT) is that the form of two and three-point correlators of quasi-primary scalar fields\footnote{Quasi-primary scalar fields (of scaling dimension $\Delta$) transform under the full $d$-dimensional conformal group as $\phi'\left(x'\right) = \left|\partial x'/\partial x\right|^{-\Delta/d}\phi\left(x\right)$. The definition of quasi-primary fields with spin is analogous.} is completely determined by conformal symmetry \cite{Polyakov70b}:
\begin{IEEEeqnarray}{l}
\left\langle\phi_1\left(x_1\right)\phi_2\left(x_2\right)\right\rangle = \left\{\begin{array}{ll} \frac{C_{12}}{x_{12}^{2\Delta}}, &\quad \Delta = \Delta_1 = \Delta_2 \\[6pt] 0, &\quad \Delta_1 \neq \Delta_2 \end{array}\right., \quad x_{ij} \equiv \left|x_i - x_j\right| \label{TwoPointFunctionsPureCFT} \\[6pt]
\left\langle\phi_1\left(x_1\right)\phi_2\left(x_2\right)\phi_3\left(x_3\right)\right\rangle = \frac{C_{123}}{x_{12}^{\Delta_1 + \Delta_2 - \Delta_3} x_{23}^{\Delta_2 + \Delta_3 - \Delta_1}x_{31}^{\Delta_3 + \Delta_1 - \Delta_2}},
\end{IEEEeqnarray}
while one-point functions vanish (except for possibly constant fields, for which $\left\langle c\right\rangle = c$)
\begin{IEEEeqnarray}{c}
\left\langle\phi_1\left(x_1\right)\right\rangle = 0,
\end{IEEEeqnarray}
where $\Delta_i$ is the scaling dimension of the scalar $\phi_i$ and $C_{12}$, $C_{123}$ are the structure constants of the corresponding correlators. If we have $n > 3$ points we may construct $n(n-3)/2$ independent conformally invariant cross (or anharmonic) ratios, e.g.\ in the case of four points,
\begin{IEEEeqnarray}{c}
\frac{x_{12}x_{34}}{x_{13}x_{24}} \quad \& \quad \frac{x_{12}x_{34}}{x_{14}x_{23}},
\end{IEEEeqnarray}
so that the corresponding $n$-point function ($n \geq 4$) has an arbitrary dependence on them, for example
\begin{IEEEeqnarray}{c}
\left\langle\phi_1\left(x_1\right)\phi_2\left(x_2\right)\phi_3\left(x_3\right)\phi_4\left(x_4\right)\right\rangle = f\left(\frac{x_{12}x_{34}}{x_{13}x_{24}}, \frac{x_{12}x_{34}}{x_{14}x_{23}}\right) \cdot \prod_{i<j}^4 x_{ij}^{\Delta/3 - \Delta_i - \Delta_j}, \qquad \Delta \equiv \sum_{i=1}^4 \Delta_i.
\end{IEEEeqnarray}
\indent Generally, quantum field theories (QFTs) can be defined even without a Lagrangian. As shown by Wightman in the 50's \cite{Wightman56}, any QFT can be reconstructed from the knowledge of its local operators $O_k\left(x\right)$ and their $n$-point correlation functions:
\begin{IEEEeqnarray}{c}
\left\langle O_1\left(x_1\right) O_2\left(x_2\right)\ldots O_n\left(x_n\right)\right\rangle. \label{nPointCorrelationFunction}
\end{IEEEeqnarray}
The latter can be determined in CFTs by means of a convergent operator product expansion (OPE) \cite{FerraraGrilloGatto73a, Polyakov74a}. E.g.\ for scalar quasi-primary fields
\begin{IEEEeqnarray}{c}
\phi_1\left(x_1\right)\phi_2\left(x_2\right) = \sum_k \frac{C_{12k}}{C_{kk}} \cdot \mathcal{P}_k\left(x_{12},\partial_2\right)\mathcal{O}_k\left(x_2\right), \label{ConformalOPE}
\end{IEEEeqnarray}
where $\mathcal{P}_k\left(x_{12},\partial_2\right)$ is a differential operator and the sum is over all the quasi-primary operators $\mathcal{O}_k$ of the CFT (with or without spin, the index contractions in \eqref{ConformalOPE} are implicit). The $(n+2)$-point correlation function can be computed recursively from the OPE \eqref{ConformalOPE}
\begin{IEEEeqnarray}{c}
\big\langle\phi_1\left(x_1\right)\phi_2\left(x_2\right)\prod_{i=3}^{n}\phi_i\left(x_i\right)\big\rangle = \sum_{k} \frac{C_{12k}}{C_{kk}} \cdot \mathcal{P}_k\left(x_{12},\partial_2\right)\big\langle\mathcal{O}_k\left(x_2\right)\prod_{i=3}^{n}\phi_i\left(x_i\right)\big\rangle
\end{IEEEeqnarray}
and so can in principle all the correlators \eqref{nPointCorrelationFunction}. It follows that CFTs are fully specified by their conformal data: $\left\{\Delta_k, \ell_k, f_k, \ldots, C_{ij} = 1, C_{ijk}\right\}$, i.e.\ their spectrum of scaling dimensions (plus some more data such as spins, flavors, etc) and their two and three-point function structure constants. Solving and constraining CFTs (by computing or severely constraining their conformal data) is the main objective of the conformal bootstrap program \cite{FerraraGrilloGatto73a, Polyakov74a}.
\subsection[Boundary and defect conformal field theories]{Boundary and defect conformal field theories \label{Subsection:DefectConformalFieldTheories}}
\begin{center}
\includegraphics[scale=0.4]{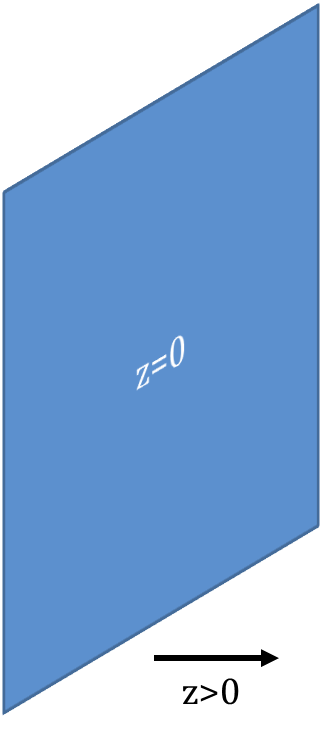}
\end{center}
\noindent Now let us consider a CFT$_d$ and introduce a plane boundary at $z = 0$, where $x_{\mu} = \left(z,\textbf{x}\right)$. The subgroup of the $d$-dimensional (Euclidean) conformal group $SO(d+1,1)$ that leaves the plane $z = 0$ invariant includes:
\begin{itemize}
\item $(d-1)$ dimensional translations: $\textbf{x}' = \textbf{x} + \textbf{a}$
\item $(d-1)$ dimensional rotations of $\textbf{x}$ (symmetric under the group $SO(d-1)$)
\item $d$ dimensional rescalings $x_{\mu}' = \alpha \cdot x_{\mu}$ and inversions $x_{\mu}' = x_{\mu}/x^2$,
\end{itemize}
where $\textbf{a}$ is a vector of the $z = 0$ plane and $\alpha$ is a constant. This is just the (Euclidean) conformal group in $d-1$ dimensions, $SO(d,1)$ \cite{Cardy84a, CardyDombLebowitz87}. The resulting setup contains a CFT$_d$ bisected by a codimension-1 boundary/interface/domain wall/defect upon which a CFT$_{d-1}$ lives; it is known as a boundary/defect conformal field theory (bCFT/dCFT).\footnote{Technically, there is a distinction between the terms boundary, interface, domain wall and defect depending on whether they accommodate extra degrees of freedom and/or boundary conditions for the bulk fields, etc. These differences need not concern us here; for all practical purposes we will be dealing with defect CFTs in the present work.} \\[6pt]
\indent Due to the presence of the $z = 0$ boundary we may form cross-ratios out of only two bulk points:
\begin{IEEEeqnarray}{c}
\xi = \frac{x_{12}^2}{\left|z_1\right|\left|z_2\right|} \qquad \& \qquad v^2 = \frac{\xi}{\xi + 1} = \frac{x_{12}^2}{x_{12}^2 + \left|z_1\right|\left|z_2\right|},
\end{IEEEeqnarray}
which implies that $n$-point bulk correlation functions are fully determined by symmetry only for $n = 1$ (i.e.\ one-point functions). The (bulk) one-point functions are nonzero only in the case of quasi-primary scalars:
\begin{IEEEeqnarray}{l}
\left\langle\phi\left(z,\textbf{x}\right)\right\rangle = \frac{C}{\left|z\right|^{\Delta}}, \label{OnePointFunctionsBoundaryCFT}
\end{IEEEeqnarray}
where $C$ is the corresponding structure constant and $\Delta$ is the scaling dimension of the scalar field $\phi$. On the other hand, $n$-point bulk correlation functions ($n \geq 2$) contain an arbitrary dependence on the cross-ratio $\xi$. E.g.\ the two-point bulk correlation function of two scalars is
\begin{IEEEeqnarray}{l}
\left\langle\phi_1\left(z_1,\textbf{x}_1\right)\phi_2\left(z_2,\textbf{x}_2\right)\right\rangle = \frac{f_{12}\left(\xi\right)}{\left|z_1\right|^{\Delta_1}\left|z_2\right|^{\Delta_2}}, \label{TwoPointFunctionsBoundaryCFT}
\end{IEEEeqnarray}
i.e.\ it does not vanish for $\Delta_1 \neq \Delta_2$. The fusion of two quasi-primary scalars in the bulk ($x_{12} \rightarrow 0$) is obviously unaffected by the presence of the defect so that the conformal OPE \eqref{ConformalOPE} can be used to determine the function $f_{12}\left(\xi\right)$ in \eqref{TwoPointFunctionsBoundaryCFT}. Using the formula \cite{McAvityOsborn95},
\begin{IEEEeqnarray}{l}
\xi^{\left(\Delta_1 + \Delta_2 - \Delta_k\right)/2} \cdot \mathcal{P}_k\left(x_{12},\partial_2\right) \frac{1}{\left|z_2\right|^{\Delta_k}} = \frac{1}{\left|z_1\right|^{\Delta_1}\left|z_2\right|^{\Delta_2}} \cdot {_2\mathcal{F}_1}\left[\frac{\Delta_k + \delta\Delta}{2},\frac{\Delta_k - \delta\Delta}{2};\Delta_k + 1 - \frac{d}{2}; -\frac{\xi}{4}\right] \qquad
\end{IEEEeqnarray}
it follows from \eqref{OnePointFunctionsBoundaryCFT}--\eqref{TwoPointFunctionsBoundaryCFT} that $f_{12}\left(\xi\right)$ is given by
\begin{IEEEeqnarray}{l}
f_{12}\left(\xi\right) = \sum_k \frac{C_{12k} C_k}{C_{kk}} \cdot \frac{1}{\xi^{\left(\Delta_1 + \Delta_2 - \Delta_k\right)/2}} \cdot {_2\mathcal{F}_1}\left[\frac{\Delta_k + \delta\Delta}{2},\frac{\Delta_k - \delta\Delta}{2};\Delta_k + 1 - \frac{d}{2}; -\frac{\xi}{4}\right], \qquad \label{ConformalBlockBulkChannel}
\end{IEEEeqnarray}
where $\xi \rightarrow 0$ and $\delta\Delta \equiv \Delta_1 - \Delta_2$. \\[6pt]
\indent Since the conformal symmetry is intact on the $z = 0$ defect, the correlation functions of boundary quasi-primary scalars $\hat{\phi}\left(\textbf{x}\right)$ satisfy the usual relations of CFT$_{(d-1)}$,
\begin{IEEEeqnarray}{l}
\big\langle\hat{\phi}_1\left(\textbf{x}_1\right)\hat{\phi}_2\left(\textbf{x}_2\right)\big\rangle = \left\{\begin{array}{ll} \frac{\hat{B}_{12}}{\textbf{x}_{12}^{2\hat{\Delta}}}, &\quad \hat{\Delta} \equiv \hat{\Delta}_1 = \hat{\Delta}_2 \\[6pt] 0, &\quad \hat{\Delta}_1 \neq \hat{\Delta}_2\end{array}\right., \quad \textbf{x}_{ij} \equiv \left|\textbf{x}_i - \textbf{x}_j\right| \\[6pt]
\big\langle\hat{\phi}_1\left(\textbf{x}_1\right)\hat{\phi}_2\left(\textbf{x}_2\right)\hat{\phi}_3\left(\textbf{x}_3\right)\big\rangle = \frac{\hat{B}_{123}}{\textbf{x}_{12}^{\hat{\Delta}_1 + \hat{\Delta}_2 - \hat{\Delta}_3} \textbf{x}_{23}^{\hat{\Delta}_2 + \hat{\Delta}_3 - \hat{\Delta}_1} \textbf{x}_{31}^{\hat{\Delta}_3 + \hat{\Delta}_1 - \hat{\Delta}_2}},
\end{IEEEeqnarray}
while all higher correlators have an explicit dependence on the boundary cross-ratios. Yet another set of dCFT data is provided by the structure constants of the bulk-boundary 2-point functions \cite{McAvityOsborn95}:
\begin{IEEEeqnarray}{l}
\big\langle\phi_1\left(z_1,\textbf{x}_1\right)\hat{\phi}_2\left(\textbf{x}_2\right)\big\rangle = \frac{B_{12}}{\left|z_1\right|^{\Delta_1 - \hat{\Delta}_2}\eta_2^{\hat{\Delta}_2}}, \qquad \eta_2 \equiv z_1^2 + \left(\textbf{x}_1 - \textbf{x}_2\right)^2.
\end{IEEEeqnarray}
The fusion of the bulk fields with the defect ($z \rightarrow 0$) is also possible. It is described by the bulk-to-defect OPE or boundary operator expansion (BOE):
\begin{IEEEeqnarray}{l}
\phi_1\left(z,\textbf{x}\right) = \sum_k \frac{B_{1k}}{\hat{B}_{kk}}\cdot\hat{\mathcal{P}}_k\big(z,\hat{\nabla}\big)\hat{\mathcal{O}}_k\left(\textbf{x}\right), \label{BoundaryOPE}
\end{IEEEeqnarray}
where the sum is over all the quasi-primary operators $\mathcal{O}_k$ of the CFT$_{(d-1)}$ on the boundary. Plugging \eqref{BoundaryOPE} into \eqref{TwoPointFunctionsBoundaryCFT} leads to a second expression for the function $f_{12}\left(\xi\right)$ \cite{McAvityOsborn95}:
\begin{IEEEeqnarray}{l}
f_{12}\left(\xi\right) = \sum_k \frac{B_{1k} B_{2k}}{\hat{B}_{kk}} \cdot \frac{1}{\xi^{\Delta_k}} \cdot {_2\mathcal{F}_1}\left[\Delta_k,\Delta_k + 1 -\frac{d}{2};2\Delta_k + 2 - d; -\frac{4}{\xi}\right]. \qquad \label{ConformalBlockDefectChannel}
\end{IEEEeqnarray}
In principle, all the correlation functions \eqref{nPointCorrelationFunction} of the defect CFT can be determined recursively from the dCFT data $\left\{\Delta_k, C_i, C_{ij}, C_{ijk}, \hat{\Delta}_k, \hat{B}_{ij}, \hat{B}_{ijk}, B_{ij}, \ldots\right\}$ by using the conformal OPE \eqref{ConformalOPE} and the BOE \eqref{BoundaryOPE}. However these data are not all independent, e.g.\ by comparing the bulk channel expression \eqref{ConformalBlockBulkChannel} for $f_{12}\left(\xi\right)$ to the one in the defect channel \eqref{ConformalBlockDefectChannel}, we obtain a number of constrains for the data. Solving defect CFTs by determining the minimal set of independent dCFT data is one of the main goals of the bootstrap program for bCFTs and dCFTs \cite{LiendoRastellivanRees12}.
\section[Holographic defect CFTs]{Holographic defect CFTs}
\noindent Superconformal field theories (SCFTs) are often holographic. The most celebrated example of a holographic SCFT is $\mathcal{N} = 4$ super Yang-Mills (SYM) theory. According to the AdS/CFT correspondence \cite{Maldacena97}, $\mathcal{N} = 4$ SYM is equivalent to type IIB superstring theory on AdS$_5\times$S$^5$:
\begin{IEEEeqnarray}{c}
\Big\{\text{Type IIB string theory on AdS}_5\times\text{S}^5\Big\} \ \Leftrightarrow \ \Big\{\mathcal{N} = 4,\ SU(N) \ \text{SYM theory in 4d}\Big\}. \qquad \label{AdS/CFTcorrespondence}
\end{IEEEeqnarray}
Holographic defect CFTs emerge on the field theory side of \eqref{AdS/CFTcorrespondence} when probe branes are inserted on the string theory side. They were introduced by Karch and Randall in an attempt to provide an explicit realization of gravity localization on an AdS$_4$ brane \cite{KarchRandall01a}.
\subsection[The D3-brane system]{The D3-brane system}
\noindent Before going on to show how a holographic defect CFT can be obtained from the AdS$_5$/CFT$_4$ correspondence \eqref{AdS/CFTcorrespondence}, let us briefly revisit an argument (originally due to Maldacena \cite{Maldacena97}) that suggests its validity. The argument examines two equivalent descriptions of a system of $N$ coincident D3-branes, namely the open and the closed string description. For more details the reader is referred to the early review \cite{MAGOO99}.
\paragraph{Open string description} Consider a system of $N$ coincident D3-branes inside the 10-dimensional Minkowski spacetime as described by type IIB string theory. The low-energy limit of this system is obtained by integrating out its massive modes. It consists of open strings that end on the D3-branes, closed strings that propagate in the 10-dimensional bulk and open-closed string interactions:
\begin{equation}
S = S_{\text{branes}} + S_{\text{bulk}} + S_{\text{interactions}}.
\end{equation}
$S_{\text{branes}}$ turns out to be $\mathcal{N} = 4$, $SU\left(N\right)$ super Yang-Mills theory in $3+1$ dimensions (plus $\alpha'$ corrections), $S_{\text{bulk}}$ is the action of type IIB supergravity in 10 dimensions (plus $\alpha'$ corrections) and $S_{\text{interactions}}$ describes the interactions between open and closed strings. To lowest order in $\alpha'$, open-closed and closed-closed string interactions are switched off and we get two decoupled systems of open and free closed strings. Schematically,
\begin{IEEEeqnarray}{c}
\left\{\begin{array}{c} \text{Open string description} \\ \text{low-energy limit} \end{array}\right\} \Rightarrow \mathcal{N} = 4,\ SU\left(N\right)\ \text{SYM}\ + \ \text{free IIB supergravity}. \label{OpenStringDescription}
\end{IEEEeqnarray}
\paragraph{Closed string description} Alternatively, the system of $N$ coincident D3-branes can be seen as a source for the fields of type IIB supergravity---the low-energy limit of type IIB string theory. The solution for the graviton field $g_{\mu\nu}$ reads
\begin{equation}
ds^2 = H^{-1/2}\left(-dt^2 + d\textbf{x}_3^2\right) + H^{1/2}\left(dr^2 + r^2 d\Omega_5^2\right), \quad H\left(r\right) \equiv 1 + \left(\frac{\ell}{r}\right)^4, \quad \ell^4 = 4\pi g_s N \alpha'^2, \label{IIBsupergravitySolution}
\end{equation}
where $g_s$ is the string coupling constant. Far from the horizon ($r \rightarrow \infty$), the above metric reduces to the 10-dimensional Minkowski metric. The near-horizon limit ($r \rightarrow 0$) of \eqref{IIBsupergravitySolution} is just AdS$_5\times\text{S}^5$ in Poincar\'{e} coordinates:
\begin{IEEEeqnarray}{c}
ds^2 = \frac{r^2}{\ell^2} \left(-dt^2 + d\textbf{x}_3^2\right) + \frac{\ell^2}{r^2} \left(dr^2 + r^2 d\Omega_5^2\right) = \frac{r^2}{\ell^2} \left(-dt^2 + d\textbf{x}_3^2\right) + \frac{\ell^2}{r^2} \sum_{i=4}^9 dx_i^2, \label{AdS5xS5metricPoincare1}
\end{IEEEeqnarray}
where $r^2 = x_4^2 + \ldots + x_9^2$. At low energies, the excitations that live far from the horizon decouple from those in the near-horizon region and the system can again be written as the sum of two decoupled systems:
\begin{IEEEeqnarray}{c}
\left\{\begin{array}{c} \text{Closed string description} \\ \text{low-energy limit} \end{array}\right\} \Rightarrow \text{IIB string theory on AdS}_5\times\text{S}^5 + \ \text{free IIB supergravity}. \qquad \label{ClosedStringDescription}
\end{IEEEeqnarray}
\paragraph{The AdS/CFT correspondence} The AdS/CFT correspondence \eqref{AdS/CFTcorrespondence} follows from the observation that the left-hand sides of \eqref{OpenStringDescription} and \eqref{ClosedStringDescription} coincide; they both describe the low-energy limit of a system of $N$ coincident D3-branes. Therefore the right-hand sides of \eqref{OpenStringDescription} and \eqref{ClosedStringDescription} should be equal; they are both expressed as a sum of two decoupled systems, one of which is free type IIB supergravity. Hence the remaining components of \eqref{OpenStringDescription} and \eqref{ClosedStringDescription}, i.e.\ $\mathcal{N} = 4,\ SU(N)$ SYM and type IIB string theory on AdS$_5\times\text{S}^5$ must be equivalent as in \eqref{AdS/CFTcorrespondence}. The rigorous formulation of \eqref{AdS/CFTcorrespondence} has the form of an equality between the corresponding partition functions:
\begin{IEEEeqnarray}{c}
\mathcal{Z}_{\text{string}}\left[\Phi\Big|_{\partial\text{AdS}}\left(x\right)\right] = \mathcal{Z}_{\text{CFT}}\left[\phi\left(x\right)\right], \label{FieldOperatorCorrespondence}
\end{IEEEeqnarray}
where $\phi$ are the sources of all the gauge-invariant operators of $\mathcal{N} = 4$ SYM and $\Phi$ are their dual fields. According to \eqref{AdS/CFTcorrespondence}, \eqref{FieldOperatorCorrespondence}, the hologram of the 10-dimensional string theory on AdS$_5\times\text{S}^5$ is $\mathcal{N} = 4$ SYM, a 4-dimensional non-abelian gauge theory that lives on the boundary of AdS$_5$ (bulk). Further, \eqref{FieldOperatorCorrespondence} implies that every boundary observable (spectra, correlation functions, scattering amplitudes, Wilson loops, etc.) possesses a dual and equal observable in the bulk. \\[6pt]
\indent The AdS/CFT correspondence \eqref{AdS/CFTcorrespondence} is a weak/strong coupling duality. Perturbative (weak-coupling) calculations of observables on one side of the correspondence are mapped to the strongly-coupled values of their dual observables. Although this property seems to hinder the attempts to check the validity of \eqref{FieldOperatorCorrespondence}, it allows to probe the strongly coupled regimes of both theories in \eqref{AdS/CFTcorrespondence}, an otherwise impossible task by means of perturbation theory. \\[6pt]
\indent The discovery of integrability on both sides of the AdS/CFT correspondence in the limit of large $N$ (planar limit) boosted our ability to compute observables beyond perturbation theory (for more see the reviews \cite{KristjansenStaudacherTseytlin09, Beisertetal12}). This enables us to solve the planar $\mathcal{N} = 4$ SYM by computing all of its observables for any value of the coupling constant $\lambda$, a huge enterprise that is still underway. The computed observables match perfectly the perturbative calculations from the (weakly-coupled) gauge theory side and the (strongly-coupled) string theory side in every case considered to date. This provides a powerful consistency check of \eqref{FieldOperatorCorrespondence} and \eqref{AdS/CFTcorrespondence}. \\[6pt]
\indent Gauge theories such as $\mathcal{N} = 4$ SYM often form the backbone of models of non-gravitational interactions. Among them, the Standard Model of elementary particles successfully unifies all of nature's forces except gravity. On the other hand, string theories have long been known to possess massless spin-2 excitations---gravitons in their spectra, while their low-energy limits are given by supergravity theories like the ones we saw above. Superstring theories are widely regarded as the ideal candidates for the description of quantum gravity. Arguably, the AdS/CFT correspondence \eqref{AdS/CFTcorrespondence} is a particle/string and a gauge/gravity duality that provides an indirect way of unifying all the fundamental interactions. \\[6pt]
\indent The catch however is that both sides of \eqref{AdS/CFTcorrespondence} are highly idealized versions of real-world theories with exact conformal symmetry ($\beta = 0$) and the maximum possible amount of supersymmetry. Real-world systems are generally characterized by finite sizes; impurities, domain walls, defects and boundaries separate regions with different properties and break many of the underlying symmetries. The introduction of a boundary/defect on both sides of the AdS/CFT correspondence \eqref{AdS/CFTcorrespondence}, not only moves us closer to real-world systems while preserving all the remarkable characteristics of AdS/CFT (e.g. holography and integrability), but it also leads to even more physical applications that include holographic models for quantum Hall systems, topological insulators and graphene.
\subsection[The D3-probe-D5 brane system]{The D3-probe-D5 brane system}
\noindent As we have just seen, IIB string theory on AdS$_5\times\text{S}^5$ is encountered very close to the system of $N$ coincident D3-branes. Now suppose that the D3-branes extend along the directions $x_1$, $x_2$, $x_3$ of \eqref{AdS5xS5metricPoincare1} and insert a single (probe) D5-brane at $x_3 = x_7 = x_8 = x_9 = 0$,
\renewcommand{\arraystretch}{1.1}
\begin{table}[H]\begin{center}\begin{tabular}{|c||c|c|c|c|c|c|c|c|c|c|}
\hline
& $t$ & $x_1$ & $x_2$ & $x_3$ & $x_4$ & $x_5$ & $x_6$ & $x_7$ & $x_8$ & $x_9$ \\ \hline
\text{D3} & $\bullet$ & $\bullet$ & $\bullet$ & $\bullet$ &&&&&& \\ \hline
\text{D5} & $\bullet$ & $\bullet$ & $\bullet$ & & $\bullet$ & $\bullet$ & $\bullet$&&& \\ \hline
\end{tabular}\caption{The D3-probe-D5 brane system.}\label{Table:D3D5system}\end{center}\vspace{-1.5em}\end{table}
\noindent There is a simple way to anticipate the worldvolume geometry of the probe D5-brane. Referring to the Poincar\'{e} metric \eqref{AdS5xS5metricPoincare1} of AdS$_5\times\text{S}^5$ we set
\begin{IEEEeqnarray}{lll}
x_4 = r \cos\psi \sin\theta \cos\varphi, \quad & x_5 = r \cos\psi \sin\theta \sin\varphi, \quad & x_6 = r \cos\psi \cos\theta \label{AdS5xS5metricCoordinates1} \\
x_7 = r \sin\psi \sin\chi \cos\zeta, \quad & x_8 = r \sin\psi \sin\chi \sin\zeta, \quad & x_9 = r \sin\psi \cos\chi, \label{AdS5xS5metricCoordinates2}
\end{IEEEeqnarray}
so that \eqref{AdS5xS5metricPoincare1} becomes
\begin{IEEEeqnarray}{c}
ds^2 = \left\{\frac{r^2}{\ell^2} \left(-dt^2 + dx_1^2 + dx_2^2 + dx_3^2\right) + \frac{\ell^2}{r^2} \, dr^2\right\} + \ell^2 \left(d\psi^2 + \cos^2\psi d\Omega_2^2 + \sin^2\psi d\Omega_2^2\right). \qquad \label{AdS5xS5metricPoincare2}
\end{IEEEeqnarray}
For $x_3 = x_7 = x_8 = x_9 = 0$, or equivalently $x_3 = \psi = 0$, \eqref{AdS5xS5metricPoincare2} becomes the metric of AdS$_4\times\text{S}^2$,
\begin{IEEEeqnarray}{c}
ds^2 = \left\{\frac{r^2}{\ell^2} \left(-dt^2 + dx_1^2 + dx_2^2\right) + \frac{\ell^2}{r^2} \, dr^2\right\} + \ell^2 d\Omega_2^2,
\end{IEEEeqnarray}
a result which is confirmed by the corresponding DBI calculation \cite{KarchRandall01b}. It can also be shown that the probe D5-brane is stable, as the potential tachyonic mode does not violate the Breitenlohner-Freedman (BF) bound.
\subsection[The D3-probe-D7 brane system]{The D3-probe-D7 brane system}
\begin{figure}[H]\begin{center}
\includegraphics[scale=0.4]{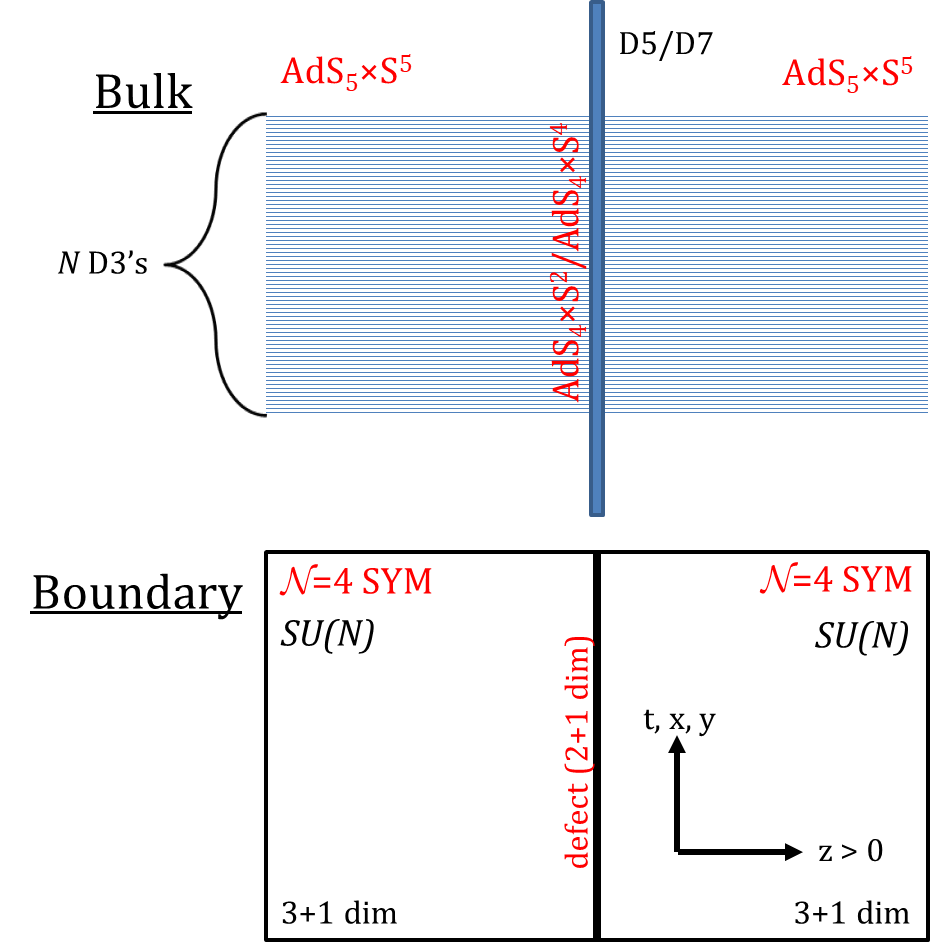} \qquad \includegraphics[scale=0.4]{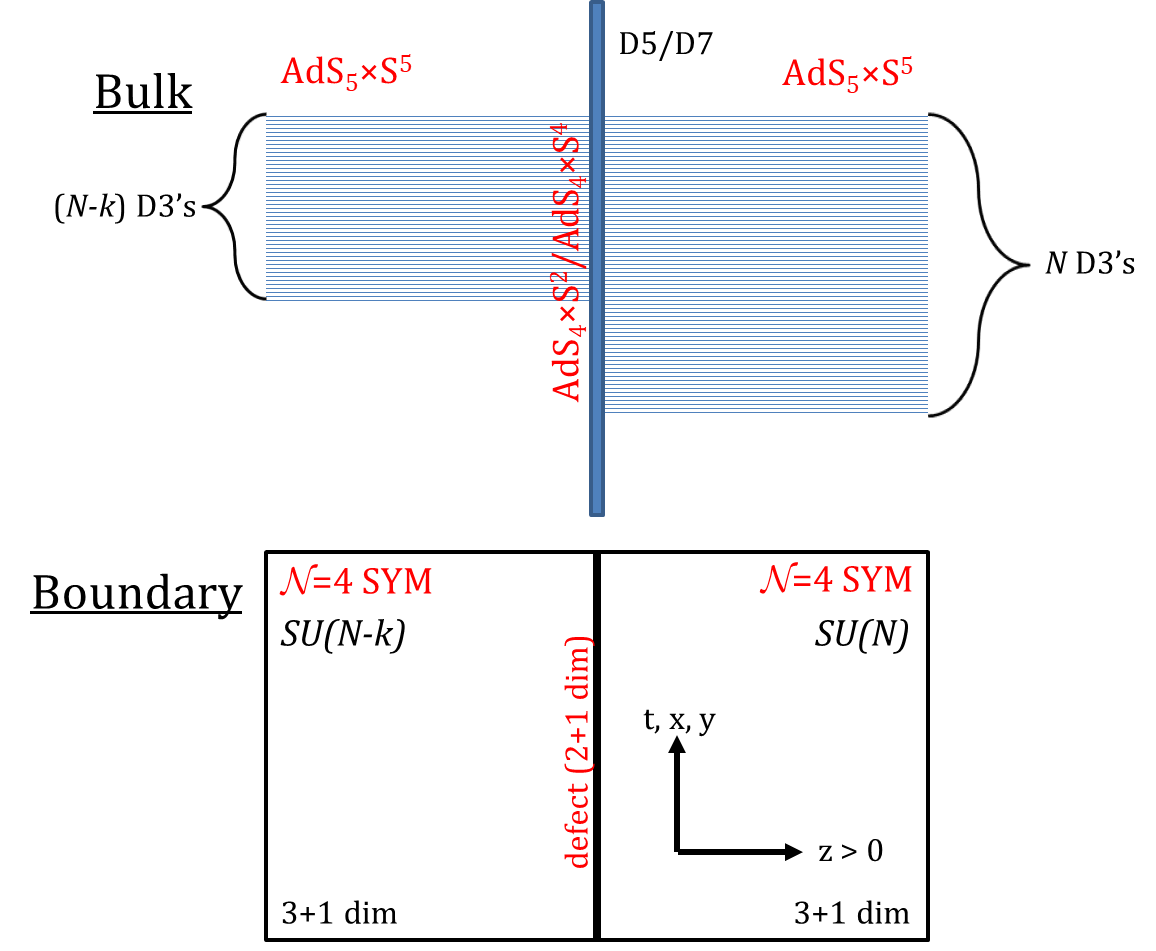}
\caption{AdS/{\color{red}d}CFT correspondence for the D3-probe-D5/D7 brane system with/without flux (right/left).} \label{Figure:D3D5D7setup}
\end{center}\vspace{-1.5em}\end{figure}

\noindent If instead of a D5-brane we place a single (probe) D7-brane at $x_3 = x_9 = 0$ as
\renewcommand{\arraystretch}{1.1}
\begin{table}[H]\begin{center}\begin{tabular}{|c||c|c|c|c|c|c|c|c|c|c|}
\hline
& $t$ & $x_1$ & $x_2$ & $x_3$ & $x_4$ & $x_5$ & $x_6$ & $x_7$ & $x_8$ & $x_9$ \\ \hline
\text{D3} & $\bullet$ & $\bullet$ & $\bullet$ & $\bullet$ &&&&&& \\ \hline
\text{D7} & $\bullet$ & $\bullet$ & $\bullet$ & & $\bullet$ & $\bullet$ & $\bullet$ & $\bullet$ & $\bullet$ & \\ \hline
\end{tabular}\caption{The D3-probe-D7 brane system.}\label{Table:D3D7system}\end{center}\vspace{-1.5em}\end{table}
\noindent the probe brane will either wrap AdS$_4\times\text{S}^4$ or AdS$_4\times\text{S}^2\times\text{S}^2$. The former geometry (the derivation of AdS$_4\times\text{S}^2\times\text{S}^2$ is similar) follows by setting $r^2 = \rho^2 + x_9^2$ in \eqref{AdS5xS5metricPoincare1},
\begin{IEEEeqnarray}{c}
ds^2 = \frac{r^2}{\ell^2} \left(-dt^2 + dx_1^2 + dx_2^2 + dx_3^2\right) + \frac{\ell^2}{r^2} \left(d\rho^2 + \rho^2 d\Omega_4^2 + dx_9^2\right) \label{AdS5xS5metricPoincare3}
\end{IEEEeqnarray}
so that $x_3 = x_9 = 0$ reduces \eqref{AdS5xS5metricPoincare3} to the line element of AdS$_4\times\text{S}^4$ :
\begin{IEEEeqnarray}{c}
ds^2 = \left\{\frac{r^2}{\ell^2} \left(-dt^2 + dx_1^2 + dx_2^2\right) + \frac{\ell^2}{r^2} \, dr^2\right\} + \ell^2 d\Omega_4^2.
\end{IEEEeqnarray}
The same result is obtained from the DBI calculation \cite{Rey08, DavisKrausShah08}. This time however there is a tachyonic mode that violates the BF bound so that a flux term will be required to stabilize the system.
\paragraph{The AdS/{\color{red}d}CFT correspondence} The insertion of the probe D5/D7-brane on the string theory side of the AdS/CFT correspondence \eqref{AdS/CFTcorrespondence} modifies it as follows (see figure \ref{Figure:D3D5D7setup}, left). The bulk is still described by IIB string theory on AdS$_5\times\text{S}^5$ which is now bisected by a probe D5/D7-brane with worldvolume geometry AdS$_4\times\text{S}^2$ (probe D5-brane) or AdS$_4\times\text{S}^4$ (probe D7-brane). The dual field theory becomes a defect CFT that consists of two interacting parts. An SCFT$_4$ ($\mathcal{N} = 4, \ SU(N)$ SYM), coupled to a CFT that lives on a flat $2 + 1$ dimensional defect \cite{DeWolfeFreedmanOoguri01, ErdmengerGuralnikKirsch02}:
\begin{IEEEeqnarray}{c}
S = S_{\mathcal{N} = 4} + S_{2+1}. \label{ActiondCFT}
\end{IEEEeqnarray}
Due to the presence of the defect, the total bosonic symmetry of the system is reduced from $SO(4,2)\times SO(6)$ to $SO(3,2)\times SO(3)\times SO(3)$ in the case of the probe D5-brane and $SO(3,2)\times SO(5)$ in the case of the probe D7-brane. In the former case (probe D5-brane) the corresponding supersymmetry is halved, i.e.\ the 32-supercharge superconformal algebra $\mathfrak{psu}\left(2,2|4\right)$ of $\mathcal{N} = 4$ SYM becomes $\mathfrak{osp}\left(4|4\right)$ which has 16 supercharges, while in the latter case (probe D7-brane) the supersymmetry is broken completely.
\indent An interesting generalization of the D3-probe-D5 brane system (table \ref{Table:D3D5system}) is obtained by supporting the worldvolume geometry of the probe D5-brane with $k$ units of magnetic flux through the S$^2$ \cite{KarchRandall01b}. The D5-brane is again stable and wraps an AdS$_4\times\text{S}^2$ inside AdS$_5\times\text{S}^5$. Its AdS coordinates satisfy
\begin{IEEEeqnarray}{c}
x_3 = \frac{\kappa\ell^2}{r}, \qquad \kappa \equiv \frac{\pi k}{\sqrt{\lambda}},
\end{IEEEeqnarray}
where $\lambda = g_{\text{YM}}^2 N$ is the 't Hooft coupling. The flux forces exactly $k$ of the $N$ D3-branes ($N \gg k$) to terminate on the D5-brane (see figure \ref{Figure:D3D5D7setup}, right). On the dual SCFT side the gauge group is different on each side of the defect, as $SU(N)\times SU(N)$ gets broken to $SU(N-k)\times SU(N)$. Equivalently, the fields of $\mathcal{N} = 4$ SYM develop nonzero vevs. Note that the zero-flux case is smoothly recovered in the $k \rightarrow 0$ limit. \\[6pt]
\indent Both probe-brane geometries of the D3-D7 system (table \ref{Table:D3D7system}) are unstable. For example, the S$^4$ component of the D7-brane that is wrapped around the equator of S$^5$ slips off towards either side of the equator after a small perturbation. One possible way to stabilize this system is by adding a homogeneous instanton on S$^4$ with an instanton number given by \cite{MyersWapler08}
\begin{IEEEeqnarray}{c}
k = d_G \equiv \frac{\left(n+1\right)\left(n+2\right)\left(n+3\right)}{6}. \label{InstantonNumber}
\end{IEEEeqnarray}
The instanton is obtained from the BPST solution by replacing the 2-dimensional irreducible representation of $SU(2)$ by an $n + 2$ dimensional one \cite{ConstableMyersTafjord01a}. The flux again forces exactly $k = d_G \ll N$ of the $N$ D3-branes to end on the D7-brane (figure \ref{Figure:D3D5D7setup}, right) and the gauge group of the dual gauge theory breaks from $SU\left(N\right)\times SU\left(N\right)$ to $SU\left(N-d_G\right)\times SU\left(N\right)$. The D7-brane is stabilized when
\begin{IEEEeqnarray}{c}
\mathfrak{Q} \equiv \frac{\pi^2}{\lambda}\left(n+1\right)\left(n+3\right) > \frac{7}{2}.
\end{IEEEeqnarray}
We note that there exist more ways to stabilize the D3-D7 system of intersecting branes (oriented as in table \ref{Table:D3D7system}). One comes about when the D7-brane is embedded in the full D3-brane geometry \eqref{IIBsupergravitySolution} instead of AdS$_5\times\text{S}^5$ \eqref{AdS5xS5metricPoincare1} in the near-horizon limit \cite{DavisKrausShah08}. Another way to lift the violation of the BF bound is by imposing an AdS cut-off $\Lambda$ \cite{KutasovLinParnachev11, MezzaliraParnachev15}. For the stabilization of the AdS$_4\times\text{S}^2\times\text{S}^2$ geometry see the paper \cite{BergmanJokelaLifschytzLippert10}.
\section[One-point functions]{One-point functions}
\begin{figure}[H]\begin{center}
\includegraphics[scale=.5]{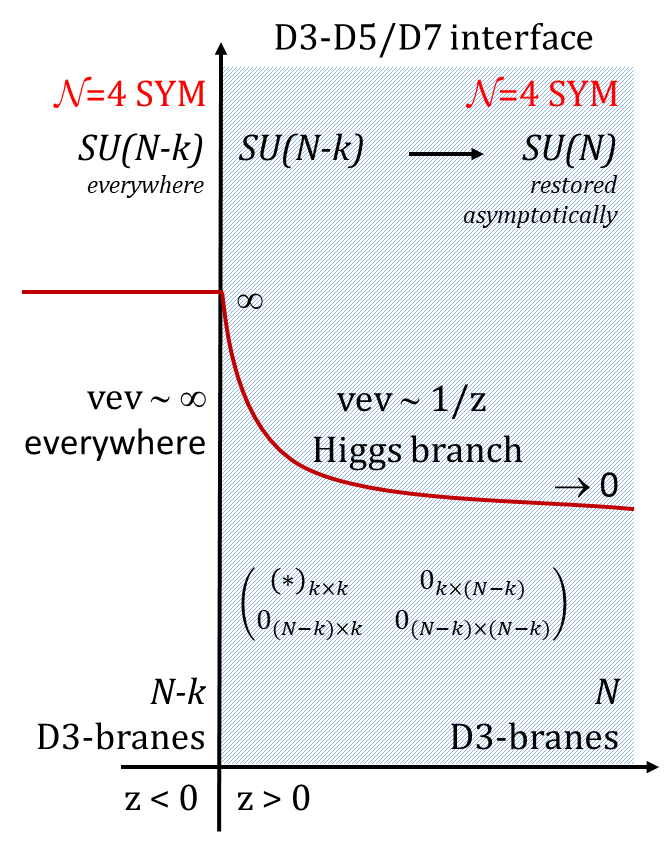}
\caption{Interfaces for the D3-D5 and the D3-D7 systems.} \label{Figure:D3D5D7interface}
\end{center}\vspace{-1.5em}\end{figure}

\noindent Once the rudiments of the AdS/{\color{red}d}CFT correspondence have been laid down for the D3-probe-D5 and the D3-probe-D7 brane systems (brane orientation in tables \ref{Table:D3D5system}--\ref{Table:D3D7system}), we need to set up a framework that will allow us to perform calculations on the defect CFT side. For this purpose we introduce the notion of the interface. Generally, an interface can be thought of as a (domain) wall that separates two different QFTs or even two copies of the same QFT. Interfaces are described by means of classical solutions of the QFT equations of motion that are known as "fuzzy-funnel" solutions \cite{ConstableMyersTafjord99, ConstableMyersTafjord01a}. \\[6pt]
\indent In the present case of the dCFTs that are dual to the D3-probe-D5 and the D3-probe-D7 brane system (tables \ref{Table:D3D5system}--\ref{Table:D3D7system}), we need an interface that separates the two copies of $\mathcal{N} = 4$ SYM with gauge groups $SU\left(N-k\right)$ (left) and $SU\left(N\right)$ (right) that lie on each side of the $2+1$ dimensional defect (see figure \ref{Figure:D3D5D7interface}). To be a valid description, the corresponding fuzzy funnel solution must preserve the global symmetry of the dCFT, i.e.\ $OSp\left(4|4\right)$ and its bosonic subgroup $SO(3,2)\times SO(3)\times SO(3)$ in the D3-D5 case and $SO(3,2)\times SO(5)$ in the D3-D7 case.
\indent For no vector and fermion fields, we want a solution of the equations of motion for the six real scalar fields of $\mathcal{N} = 4$ SYM ($\mu = 0,\ldots,3, \ \text{a} = 1,\ldots,4$):
\begin{IEEEeqnarray}{c}
A_{\mu} = \psi_{\text{a}} = 0, \quad \& \quad \frac{d^2\phi_i}{dz^2} = \left[\phi_j, \left[\phi_j, \phi_i\right]\right], \quad i,j = 1,\ldots,6. \label{SYMequations}
\end{IEEEeqnarray}
A manifestly $SO(3) \times SO(3) \simeq SU(2) \times SU(2)$ symmetric solution of \eqref{SYMequations} for $z>0$, is \cite{Diaconescu96, GiveonKutasov98, ConstableMyersTafjord99}:
\begin{IEEEeqnarray}{l}
\phi_{2i-1}\left(z\right) = \frac{1}{z}\left[\begin{array}{cc} \left(t_i\right)_{k\times k} & 0_{k\times \left(N - k\right)} \\ 0_{\left(N - k\right)\times k} & 0_{\left(N - k\right)\times \left(N - k\right)} \end{array}\right] \quad \& \quad \phi_{2i} = 0, \qquad i = 1,2,3, \label{FuzzyFunnelSU2}
\end{IEEEeqnarray}
where the matrices $t_i$ furnish a $k$-dimensional representation of $\mathfrak{so}\left(3\right) \simeq \mathfrak{su}\left(2\right)$,
\begin{IEEEeqnarray}{c}
\left[t_i, t_j\right] = i \epsilon_{ijk}t_k. \label{SU2algebra}
\end{IEEEeqnarray}
The fuzzy-funnel solution \eqref{FuzzyFunnelSU2} that describes the D3-probe-D5 brane interface also satisfies the Nahm equations:
\begin{IEEEeqnarray}{l}
\frac{d\phi_i}{dz} = \frac{i}{2} \epsilon_{ijk}\left[\phi_j, \phi_k\right]
\end{IEEEeqnarray}
and therefore induces a set of boundary conditions for the scalars that preserves one-half of the original supersymmetry \cite{GaiottoWitten08a}, as it should. The D3-probe-D7 brane interface is described by the following $SO(5)$ symmetric solution of \eqref{SYMequations} \cite{ConstableMyersTafjord01a}:
\begin{IEEEeqnarray}{l}
\phi_{i}\left(z\right) = \frac{1}{\sqrt{8} \, z} \cdot \left[\begin{array}{cc} \left(G_i\right)_{d_{G}\times d_{G}} & 0_{d_{G}\times \left(N - d_{G}\right)} \\ 0_{\left(N - d_{G}\right)\times d_{G}} & 0_{\left(N - d_{G}\right)\times \left(N - d_{G}\right)} \end{array}\right] \quad \& \quad \phi_{6} = 0, \qquad i = 1,\ldots,5, \label{FuzzyFunnelSO5}
\end{IEEEeqnarray}
where $G_i$ are the $d_G$-dimensional fuzzy S$^4$ matrices \cite{CastelinoLeeTaylor97, ConstableMyersTafjord01a} and $k = d_G$ is the instanton number \eqref{InstantonNumber}. The commutators $G_{ij}$ of the $G$-matrices furnish a $d_G$-dimensional (anti-hermitian) irreducible representation of $\mathfrak{so}\left(5\right)\simeq\mathfrak{sp}\left(4\right)$,
\begin{IEEEeqnarray}{lc}
\left[G_{ij},G_{kl}\right] = 2\left(\delta_{jk}G_{il} + \delta_{il}G_{jk} - \delta_{ik}G_{jl} - \delta_{jl}G_{ik}\right), \qquad G_{ij} \equiv \frac{1}{2}\left[G_i,G_j\right]. \label{SO5algebra}
\end{IEEEeqnarray}
\subsection[Overlaps of matrix product states]{Overlaps of matrix product states \label{Subsection:MatrixProductStateOverlaps}}
\noindent As it has already been explained in \S\ref{Subsection:DefectConformalFieldTheories}, the insertion of a plane codimension-1 boundary in the bulk of a $d$-dimensional CFT fixes the form of one-point functions of local scalar operators. Only the one-point functions of quasi-primary scalars are non-zero in the bulk and are given by \eqref{OnePointFunctionsBoundaryCFT}:
\begin{IEEEeqnarray}{l}
\left\langle O\left(z,\textbf{x}\right)\right\rangle = \frac{C}{z^{\Delta}}, \qquad z>0, \label{OnePointFunctionsDefectCFTs}
\end{IEEEeqnarray}
where $\Delta$ is the scaling dimension of the operator $O\left(z,\textbf{x}\right)$ and $C$ its one-point function structure constant. Following Nagasaki and Yamaguchi \cite{NagasakiYamaguchi12}, the tree-level one-point functions of single-trace scalar operators of a dCFT can be calculated in the planar limit ($N \rightarrow \infty$) from the corresponding fuzzy-funnel solution. E.g.\ for the dCFT that is dual to the D3-probe-D5 brane system,
\begin{IEEEeqnarray}{c}
O\left(z,\textbf{x}\right) = \Psi^{i_1 \ldots i_L}\text{tr}\left[\phi_{i_1}\ldots\phi_{i_L}\right] \ \xrightarrow{ \ \eqref{FuzzyFunnelSU2} \ } \ \frac{1}{z^L} \cdot \Psi^{i_1 \ldots i_L}\text{tr}\left[t_{i_1}\ldots t_{i_L}\right] \label{SingleTraceScalarOperators}
\end{IEEEeqnarray}
and similarly for the dCFT that is dual to the D3-probe-D7 brane system. Generally, $\Psi_{i_1 \ldots i_L}$ is an $SO\left(6\right)$ symmetric tensor, invariant under cyclic permutations. The one-point function structure constant $C$ of \eqref{OnePointFunctionsDefectCFTs} is then given by
\begin{IEEEeqnarray}{c}
C = \frac{1}{\sqrt{L}}\left(\frac{8\pi^2}{\lambda}\right)^{L/2} \cdot \frac{\left\langle\text{MPS}|\Psi\right\rangle}{\left\langle\Psi|\Psi\right\rangle^{1/2}}, \qquad \left\langle\Psi|\Psi\right\rangle \equiv \Psi^{i_1 \ldots i_L} \Psi_{i_1 \ldots i_L}, \label{OnePointFunctionStructureConstant}
\end{IEEEeqnarray}
where $\left\langle\text{MPS}|\Psi\right\rangle$ is the overlap of the matrix product state (MPS) with the state $\left|\Psi\right\rangle$. For the two cases of interest \eqref{FuzzyFunnelSU2}, \eqref{FuzzyFunnelSO5}
\begin{IEEEeqnarray}{c}
\langle\text{MPS}|_{\text{D3-D5}} = \text{tr}_{a} \prod_{l=1}^L \sum_{i=1}^3 \langle\phi_{2i-1}| \otimes t_i, \qquad \langle\text{MPS}|_{\text{D3-D7}} = \text{tr}_{a} \prod_{l=1}^L \sum_{i=1}^5 \langle\phi_{i}| \otimes G_i, \label{MatrixProductStates}
\end{IEEEeqnarray}
where the index $a$ refers to the auxiliary space of the color indices of the matrices $t_i$ and $G_i$ on which the trace operation applies. At tree-level, \eqref{MatrixProductStates} implies
\begin{IEEEeqnarray}{c}
\left\langle\text{MPS}|\Psi\right\rangle_{\text{D3-D5}} = \Psi^{i_1 \ldots i_L}\text{tr}\left[t_{i_1}\ldots t_{i_L}\right], \qquad \left\langle\text{MPS}|\Psi\right\rangle_{\text{D3-D7}} = \Psi^{i_1 \ldots i_L}\text{tr}\left[G_{i_1}\ldots G_{i_L}\right]. \label{MatrixProductStateOverlaps}
\end{IEEEeqnarray}
The overall normalization factor of the structure constant \eqref{OnePointFunctionStructureConstant} originates from the normalization of single-trace operators as $O \rightarrow (2\pi)^L\cdot O/(\lambda^{L/2}\sqrt{L})$, which ensures that their two-point functions \eqref{TwoPointFunctionsPureCFT} will be normalized to unity:
\begin{IEEEeqnarray}{c}
\left\langle O\left(x_1\right) O\left(x_2\right)\right\rangle = \frac{1}{\left|x_1 - x_2\right|^{2\Delta}},
\end{IEEEeqnarray}
within $\mathcal{N} = 4$ SYM, i.e.\ without the defect \cite{LeeMinwallaRangamaniSeiberg98}.
\paragraph{Chiral primary operators} The one-point functions of the chiral primary operators (CPOs)
\begin{IEEEeqnarray}{c}
O_{\text{CPO}}\left(x\right) = \frac{1}{\sqrt{L}}\left(\frac{8\pi^2}{\lambda}\right)^{L/2} \cdot C^{i_1 \ldots i_L}\text{Tr}\left[\phi_{i_1}\ldots\phi_{i_L}\right], \label{ChiralPrimaryOperators}
\end{IEEEeqnarray}
where $C_{i_1 \ldots i_L}$ are symmetric traceless tensors that are associated to the S$^5$ spherical harmonics,
\begin{IEEEeqnarray}{c}
Y_L = C^{i_1 \ldots i_L} \hat{x}_{i_1}\ldots\hat{x}_{i_L}, \qquad C^{i_1 \ldots i_L}C_{i_1 \ldots i_L} = 1, \qquad \hat{x}_{i-3} \equiv \frac{x_{i}}{r}, \qquad \sum_{i=4}^9 x_{i}^2 = r, \label{SphericalHarmonics}
\end{IEEEeqnarray}
have been calculated at tree-level for the matrix product states \eqref{MatrixProductStates}, in the planar limit \cite{NagasakiYamaguchi12, KristjansenSemenoffYoung12b}. In the D3-D5 case \cite{NagasakiYamaguchi12}, the spherical harmonics \eqref{SphericalHarmonics} are invariant under the global symmetry $SO(3)\times SO(3) \subseteq SO(6)$ of the interface \eqref{FuzzyFunnelSU2}, so that the coordinates $x_i$ are given by \eqref{AdS5xS5metricCoordinates1}--\eqref{AdS5xS5metricCoordinates2}:
\begin{IEEEeqnarray}{c}
\sum_{i=1}^3 \hat{x}_{i}^2 = \cos^2\psi, \qquad \sum_{i=4}^6 \hat{x}_{i}^2 = \sin^2\psi
\end{IEEEeqnarray}
and correspond to the metric \eqref{AdS5xS5metricPoincare2}. For small values of the 't Hooft coupling constant $\lambda$ (weak coupling), the one-point functions read
\begin{IEEEeqnarray}{c}
\left\langle O_{\text{CPO}}\left(x\right)\right\rangle = \left(k^2 - 1\right)^{L/2} \cdot \frac{k }{\sqrt{L}} \left(\frac{2 \pi^2}{\lambda}\right)^{L/2} \frac{Y_{L}\left(\pi/2\right)}{z^L}, \qquad k \ll N \rightarrow \infty. \label{OnePointFunctionsCPOweakD3D5}
\end{IEEEeqnarray}
The one-point functions of the CPO's \eqref{ChiralPrimaryOperators} can also be computed at strong coupling ($N, \lambda \rightarrow \infty$) for the D3-D5 system (table \ref{Table:D3D5system}). The supergravity calculation gives
\begin{IEEEeqnarray}{c}
\left\langle O_{\text{CPO}}\left(x\right)\right\rangle = \frac{k^{L+1}}{\sqrt{L}} \left(\frac{2\pi^2}{\lambda}\right)^{L/2} \frac{Y_{L}\left(\pi/2\right)}{z^L} \cdot \left[1 + \frac{\lambda \, \textrm{I}_1}{\pi^2 k^2} + \ldots\right], \quad \textrm{I}_1 \equiv \frac{3}{2} + \frac{\left(L - 2\right)\left(L - 3\right)}{4\left(L - 1\right)}. \label{OnePointFunctionsCPOstrongD3D5} \qquad
\end{IEEEeqnarray}
To lowest-order in $\lambda$, \eqref{OnePointFunctionsCPOstrongD3D5} agrees with the weak-coupling result \eqref{OnePointFunctionsCPOweakD3D5} in the double-scaling limit $k, \lambda \rightarrow \infty$, $k^2/\lambda = $ const. \\[6pt]
\indent For the D3-D7 interface \eqref{FuzzyFunnelSO5}, the computation of the tree-level one-point functions of the CPOs \eqref{ChiralPrimaryOperators} is similar \cite{KristjansenSemenoffYoung12b}. If $Y_{L}\left(\theta\right)$ is the $SO(5)$ spherical harmonic ($Y_{L = \text{odd}}\left(0\right) = 0$), then
\begin{IEEEeqnarray}{c}
\left\langle O_{\text{CPO}}\left(x\right)\right\rangle = \frac{d_G}{\sqrt{L}} \left(\frac{\pi^2 c_{\textsc{\tiny G}}}{\lambda}\right)^{L/2} \frac{Y_{L}\left(0\right)}{z^L}, \qquad c_{\textsc{\tiny G}} \equiv n\left(n+4\right), \qquad d_G \ll N \rightarrow \infty, \label{OnePointFunctionsCPOweakD3D7}
\end{IEEEeqnarray}
gives the tree-level one-point functions of \eqref{ChiralPrimaryOperators} for the D3-D7 MPS. At strong coupling ($\lambda \rightarrow \infty$),
\begin{IEEEeqnarray}{c}
\left\langle O_{\text{CPO}}\left(x\right)\right\rangle = \frac{n^3}{\sqrt{L}} \left(\frac{\pi^2 n^2}{\lambda}\right)^{L/2} \frac{Y_{L}\left(0\right)}{z^L}, \qquad N \rightarrow \infty,
\end{IEEEeqnarray}
which again agrees with the weak-coupling result \eqref{OnePointFunctionsCPOweakD3D7} in the limit $n, \lambda \rightarrow \infty$, $n^2/\lambda = $ const.
\paragraph{Highest-weight states} Single-trace scalar operators like \eqref{SingleTraceScalarOperators} have definite scaling dimensions $\Delta$ and so they are eigenstates of a dilatation operator. To lowest order in $\lambda$, this dilatation operator coincides with the integrable $\mathfrak{so}\left(6\right)$ closed (i.e.\ with periodic boundary conditions) spin chain:
\begin{IEEEeqnarray}{c}
\mathbb{D} = L \cdot \mathbb{I} + \frac{\lambda}{8\pi^2}\cdot \mathbb{H} + \sum_{n = 2}^\infty \lambda^n\cdot\mathbb{D}_n, \qquad \mathbb{H} = \sum_{j = 1}^L \left(\mathbb{I}_{j,j + 1} - \mathbb{P}_{j,j + 1} + \frac{1}{2} \, \mathbb{K}_{j,j + 1}\right), \label{DilatationOperator}
\end{IEEEeqnarray}
that describes the mixing of single-trace scalar operators up to one loop in $\mathcal{N} = 4$ SYM \cite{MinahanZarembo03}. The identity, permutation and trace operators that appear in \eqref{DilatationOperator} are respectively given by:
\begin{IEEEeqnarray}{l}
\mathbb{I} \cdot \left|\ldots\phi_a\phi_b\ldots\right> = \left|\ldots\phi_a\phi_b\ldots\right> \label{IdentityOperator} \\[6pt]
\mathbb{P} \cdot \left|\ldots\phi_a\phi_b\ldots\right> = \left|\ldots\phi_b\phi_a\ldots\right> \label{PermutationOperator} \\
\mathbb{K} \cdot \left|\ldots\phi_a\phi_b\ldots\right> = \delta_{ab} \, \sum_{c = 1}^6\left|\ldots\phi_c\phi_c\ldots\right>. \label{TraceOperator}
\end{IEEEeqnarray}
\noindent In what follows, we will only consider the 1-point functions of bulk operators that correspond to highest-weight eigenstates of the periodic spin chain \eqref{DilatationOperator}, i.e.\ the $\left|\Psi\right\rangle$'s in \eqref{OnePointFunctionStructureConstant} are constructed with the Bethe ansatz. For a study of 1-point functions of D3-D5 descendant operators, see \cite{deLeeuwIpsenKristjansenVardinghusWilhelm17}.
\subsection[Coordinate Bethe ansatz]{Coordinate Bethe ansatz}
\noindent The eigenstates $|\Psi\rangle$ of the $\mathfrak{so}\left(6\right)$ spin chain \eqref{DilatationOperator} are constructed by means of the coordinate nested Bethe ansatz (cNBA).\footnote{For pedagogical introductions to the nested coordinate Bethe ansatz, the reader is referred to the theses \cite{Okamura08, Escobedo12}.} Nested Bethe states have the form:
\begin{IEEEeqnarray}{c}
\left|\Psi\right\rangle = \sum_{x_j} \psi\left(\textbf{u}_1, \textbf{u}_2, \textbf{u}_3\right) \cdot \left|\left\{n_{1,j}\right\},\left\{n_{2,k}\right\},\left\{n_{3,l}\right\}\right\rangle, \label{NestedBetheEigenstates}
\end{IEEEeqnarray}
where $\textbf{u}_{1,2,3}$ are three sets of rapidities for the spin chain excitations at positions $x_j$,
\begin{IEEEeqnarray}{c}
\textbf{u}_{a} = \left\{u_{a,i}\right\}, \quad a = 1,2,3, \quad i = 1,\ldots, N_{a}, \qquad u_{1,j} = \frac{1}{2}\cot\frac{p_j}{2}
\end{IEEEeqnarray}
and $p_j$ are the momenta of the level-1 excitations $n_{1,j}$ ($j = 1,\ldots, N_{1}$).
\paragraph{Nesting kets} The three sets or levels of spin chain excitations $n_{a,i}$ in \eqref{NestedBetheEigenstates} are associated with the nodes of the $\mathfrak{so}\left(6\right)$ Dynkin diagram:
\begin{center}
\begin{tikzpicture}
\draw[fill=black]

(8.6,0) circle [radius=.08]

(8.6,0) node [left] {$N_1$}

(8.6,0) --++ (30:1)
      circle [radius=.08] node [right] {$N_2$}

(8.6,0) --++ (-30:1)
      circle [radius=.08] node [right] {$N_3$}
;

\end{tikzpicture}
\end{center}
The nesting procedure for an $\mathfrak{so}\left(6\right)$ spin chain of length $L$, involves exciting the middle (momentum-carrying) node of the Dynkin diagram by inserting $N_1$ (level-1) excitations at the positions $n_{1,j} = x_j$ of the spin chain. Then the other two nodes of the Dynkin diagram are excited by placing $N_2$ (level-2) and $N_3$ (level-3) excitations at the positions $n_{2,k}$ and $n_{3,l}$ of the reduced chain (i.e.\ the spin chain that is formed by the level-1 excitations) respectively. We are led to the following nesting ket:
\begin{IEEEeqnarray}{c}
\left|\left\{n_{1,j}\right\},\left\{n_{2,k}\right\},\left\{n_{3,l}\right\}\right\rangle, \label{NestingKet}
\end{IEEEeqnarray}
where
\begin{IEEEeqnarray}{ll}
1 \leq n_{1,1}\leq \ldots \leq n_{1,N_1} \leq L, \qquad & j = 1,\ldots,N_1 \\
1 \leq n_{2,1}\leq \ldots \leq n_{2,N_2} \leq N_1, & k = 1,\ldots,N_2 \\
1 \leq n_{3,1}\leq \ldots \leq n_{3,N_3} \leq N_1, & l = 1,\ldots,N_3.
\end{IEEEeqnarray}
The total weight of the representation is
\begin{IEEEeqnarray}{c}
\omega = L q - N_1 \alpha_1 - N_2 \alpha_2 - N_3 \alpha_3 = \left(L - N_1, N_1 - N_2 - N_3, N_2 - N_3\right) = \left(J_1, J_2, J_3\right), \qquad \label{RepresentationSO6}
\end{IEEEeqnarray}
where $J_{1,2,3}$ are the corresponding Cartan charges. In \eqref{RepresentationSO6}, $q = \left(1,0,0\right)$ is the highest weight of $\mathfrak{so}\left(6\right)$ and $\alpha_1 = \left(1,-1,0\right)$, $\alpha_2 = \left(0,1,-1\right)$, $\alpha_3 = \left(0,1,1\right)$ its simple roots so that the $\mathfrak{so}(6)$ Cartan matrix is given by:
\begin{IEEEeqnarray}{c}
M_{ab} = \frac{2\alpha_{a}\cdot \alpha_{b}}{\alpha_b^2} = \left(\begin{array}{ccc} 2 & -1 & -1 \\ -1 & 2 & 0 \\ -1 & 0 & 2 \end{array}\right), \qquad q_a = \left(\begin{array}{c} 1 \\ 0 \\ 0 \end{array}\right), \label{CartanMatrixSO6}
\end{IEEEeqnarray}
where the indices $a,b$ run from 1 to 3. The Dynkin indices of the representation \eqref{RepresentationSO6} are
\begin{IEEEeqnarray}{ll}
\left[\omega \cdot \alpha_2,\omega \cdot \alpha_1,\omega \cdot \alpha_3\right] = \left[N_1 - 2N_2, L - 2N_1 + N_2, N_1 - 2N_3\right] = \left[J_2 - J_3, J_1 - J_2, J_2 + J_3\right]. \qquad
\end{IEEEeqnarray}
The next step is to replace each nesting ket that shows up in \eqref{NestedBetheEigenstates} by a single-trace scalar operator like \eqref{SingleTraceScalarOperators}. To this end, the spin chain excitations of the nesting kets \eqref{NestingKet} must be mapped to the scalar fields of the dCFT. It is customary to combine the six real scalar fields $\phi_i$ ($i = 1,\ldots,6$) into three complex scalars as follows:
\begin{IEEEeqnarray}{lll}
W = \phi_1 + i\phi_2, \qquad \qquad &Y = \phi_3 + i\phi_4, \qquad \qquad &Z = \phi_5 + i\phi_6 \label{HolomorphicFields} \\[6pt]
\bar{W} = \phi_1 - i\phi_2, \qquad \qquad &\bar{Y} = \phi_3 - i\phi_4, \qquad \qquad &\bar{Z} = \phi_5 - i\phi_6. \label{AntiHolomorphicFields}
\end{IEEEeqnarray}
Now the field content of the representation \eqref{RepresentationSO6} is determined from the following associations:
\begin{IEEEeqnarray}{lll}
Z \sim q, \qquad & W \sim q - \alpha_1, \qquad & Y \sim q - \alpha_1 - \alpha_2 \qquad \label{HolomporphicFieldExcitations} \\[6pt]
\bar{Z} \sim q - 2\alpha_1 - \alpha_2 - \alpha_3, \qquad & \bar{W} \sim q - \alpha_1 - \alpha_2 - \alpha_3, \qquad & \bar{Y} \sim q - \alpha_1 - \alpha_3, \qquad \label{AntiholomporphicFieldExcitations}
\end{IEEEeqnarray}
where obviously the complex scalar field $Z$ corresponds to the vacuum state of the spin chain, the field $W$ to the excitations of level-1 , $Y$ to level-2, while the excitations of the third level are the three anti-holomorphic fields $\bar{Z}$, $\bar{W}$, $\bar{Y}$.
\paragraph{Nested Bethe wavefunction} The other basic ingredient of the $\mathfrak{so}\left(6\right)$ Bethe eigenstates $\left|\Psi\right\rangle$ in \eqref{NestedBetheEigenstates} is the nested Bethe wavefunction $\psi\left(\textbf{u}_1, \textbf{u}_2, \textbf{u}_3\right)$. Omitting certain subtleties concerning the contribution of the anti-holomorphic excitations $\bar{W}$ and $\bar{Z}$,\footnote{Details can be found in appendix E.5 of \cite{BassoCoronadoKomatsuLamVieiraZhong17}.} the wavefunction $\psi$ is given by:
\begin{IEEEeqnarray}{l}
\psi\left(\textbf{u}_1, \textbf{u}_2, \textbf{u}_3\right) = \sum_{\sigma_1} A_1\left(\sigma_1\right) \cdot \prod_{j = 1}^{N_1} \left(\frac{u_{1,\sigma_{1,j}} + \frac{i}{2}}{u_{1,\sigma_{1,j}} - \frac{i}{2}}\right)^{n_{1,j}} \times \psi_2\left(\sigma_1\right) \times \psi_3\left(\sigma_1\right), \qquad \qquad \label{NestedWavefunction1} \\[6pt]
\psi_a\left(\sigma_1\right) = \sum_{\sigma_a} A_a\left(\sigma_a\right) \cdot \prod_{k = 1}^{N_a} \prod_{r = 1}^{n_{a,k}}\frac{\left(u_{a,\sigma_{a,k}} - u_{1,\sigma_{1,r}} + \frac{i}{2}\right)^{\delta_{r \neq n_{a,k}}}}{u_{a,\sigma_{a,k}} - u_{1,\sigma_{1,r}} - \frac{i}{2}}, \qquad a = 2,3, \qquad \label{NestedWavefunction2}
\end{IEEEeqnarray}
where $u_{a,j}$ is the rapidity of the level-$a$ excitation at $n_{a,j}$ and $\sigma_a$ denotes the permutations of the set $\left(1,2,\ldots,N_a\right)$ for $a = 1,2,3$. The coefficients $A_a\left(\sigma_a\right)$ are determined from the following formula,
\begin{IEEEeqnarray}{l}
A_a\left(\ldots,j,k,\ldots\right) = S_a\left(u_{a,j}, u_{a,k}\right)\cdot A_a\left(\ldots,k,j,\ldots\right),
\end{IEEEeqnarray}
where $A_a\left(1,2,\ldots N_a\right) = 1$ and $S_a$ is the S-matrix for the Dynkin node $a$,
\begin{IEEEeqnarray}{l}
S_a\left(u_{a,j}, u_{a,k}\right) = \frac{u_{a,j} - u_{a,k} + i}{u_{a,j} - u_{a,k} - i}\,.
\end{IEEEeqnarray}
\paragraph{Bethe roots} The Bethe wavefunction $\psi$ must be periodic at each nesting level. Imposing its invariance under shifts of $n_{a,j}$ leads to the following set of Bethe equations:
\begin{IEEEeqnarray}{rl}
\left(\frac{u_{1,i} + \frac{i}{2}}{u_{1,i} - \frac{i}{2}}\right)^L &= \prod_{{\substack{j = 1 \\[2pt] j \neq i}}}^{N_1}\frac{u_{1,i} - u_{1,j} + i}{u_{1,i} - u_{1,j} - i}\prod_{k = 1}^{N_2}\frac{u_{1,i} - u_{2,k} - \frac{i}{2}}{u_{1,i} - u_{2,k} + \frac{i}{2}}\prod_{l = 1}^{N_3}\frac{u_{1,i} - u_{3,l} - \frac{i}{2}}{u_{1,i} - u_{3,l} + \frac{i}{2}} \label{BetheEquations1} \\[6pt]
1 &= \prod_{j = 1}^{N_1}\frac{u_{2,i} - u_{1,j} - \frac{i}{2}}{u_{2,i} - u_{1,j} + \frac{i}{2}}\prod_{{\substack{k =1 \\[2pt] k \neq i}}}^{N_2}\frac{u_{2,i} - u_{2,k} + i}{u_{2,i} - u_{2,k} - i} \label{BetheEquations2} \\[6pt]
1 &= \prod_{j = 1}^{N_1}\frac{u_{3,i} - u_{1,j} - \frac{i}{2}}{u_{3,i} - u_{1,j} + \frac{i}{2}}\prod_{{\substack{l =1 \\[2pt] l \neq i}}}^{N_3}\frac{u_{3,i} - u_{3,l} + i}{u_{3,i} - u_{3,l} - i}\,, \label{BetheEquations3}
\end{IEEEeqnarray}
which are satisfied by the rapidities of the excitations or Bethe roots. The cyclicity of the trace in single-trace operators such as \eqref{SingleTraceScalarOperators} also implies that the total momentum of level-1 excitations (or momentum-carrying Bethe roots) is always zero:
\begin{IEEEeqnarray}{c}
\prod_{i=1}^{N_1}\frac{u_{1,i}+\frac{i}{2}}{u_{1,i}-\frac{i}{2}} = 1 \ \Leftrightarrow \ \sum^{N_1}_{i=1} p_{1,i} = 0 \qquad \left(\text{momentum conservation}\right). \label{MomentumConservation}
\end{IEEEeqnarray}
It can be proven that the Bethe states $\left|\Psi\right\rangle$ constructed via \eqref{NestedBetheEigenstates} and mapped to a single-trace operator via \eqref{HolomporphicFieldExcitations}--\eqref{AntiholomporphicFieldExcitations}, are indeed eigenstates of the $\mathfrak{so}\left(6\right)$ spin chain \eqref{DilatationOperator} if the rapidities $u_{a,j}$ solve the Bethe equations \eqref{BetheEquations1}--\eqref{BetheEquations3} and obey \eqref{MomentumConservation}. The corresponding eigenvalues/operator scaling dimensions $\Delta$ are given by:
\begin{IEEEeqnarray}{c}
\Delta = L + \frac{\lambda \, \gamma}{8\pi^2} + \mathcal{O}\left(\lambda^2\right), \qquad \gamma \equiv \sum^{N_1}_{j=1} \frac{1}{u_{1,j}^2 + \frac{1}{4}}. \label{BetheStateEnergy}
\end{IEEEeqnarray}
The Bethe wavefunction $\psi$ in \eqref{NestedWavefunction1}--\eqref{NestedWavefunction2} simplifies significantly if one instead of the full $SO(6)$ sector which includes all the fields \eqref{HolomorphicFields}--\eqref{AntiHolomorphicFields}, considers its two subsectors $SU(2)$ and $SU(3)$.
\paragraph{The $SU\left(2\right)$ sector} The $SU(2)$ subsector contains only the two holomorphic scalars $Z \sim \left|\downarrow\right\rangle$ and $W \sim \left|\uparrow\right\rangle$. The trace operator \eqref{TraceOperator} acts only on anti-holomorphic fields and so it does not contribute to the mixing matrix \eqref{DilatationOperator} which reduces to the Hamiltonian of the Heisenberg XXX$_{1/2}$ spin chain:
\begin{IEEEeqnarray}{c}
\mathbb{H} = \sum_{j = 1}^L \left(\mathbb{I}_{j,j + 1} - \mathbb{P}_{j,j + 1}\right). \label{HeisenbergXXXonehalf}
\end{IEEEeqnarray}
The Bethe eigenstate $\left|\Psi\right\rangle = \left|\textbf{p}\right\rangle$ becomes:
\begin{IEEEeqnarray}{c}
\left|\textbf{p}\right\rangle = \mathfrak{N} \cdot \sum_{\sigma \in S_M}\sum_{1\leq x_1 \leq \ldots \leq x_M \leq L} \exp\left[i\sum_{k}p_{\sigma_k}x_k + \frac{i}{2} \sum_{j<k} \theta_{\sigma_j\sigma_k}\right] \left|\textbf{x}\right\rangle, \ \left|\textbf{p}\right\rangle\equiv\left|p_1,p_2,\ldots,p_{M}\right\rangle, \qquad \label{SU2wavefunction}
\end{IEEEeqnarray}
where $M = N_1$ denotes the number of magnons and
\begin{IEEEeqnarray}{c}
\left|\textbf{x}\right\rangle \equiv \left|x_1,x_2,\ldots,x_M\right\rangle \equiv |\downarrow \ldots \downarrow{\color{red}\underset{x_1}{\uparrow}}\downarrow \ldots \downarrow{\color{red}\underset{x_2}{\uparrow}}\downarrow{\color{red} \ldots} \downarrow{\color{red}\underset{x_M}{\uparrow}}\downarrow \ldots \downarrow\rangle = S_{x_1}^{+} \ldots S_{x_M}^{+}\left|0\right\rangle.
\end{IEEEeqnarray}
The vacuum state $\left|0\right\rangle$ and the raising and lowering operators $S^{\pm}$ have been defined as
\begin{IEEEeqnarray}{c}
\left|0\right\rangle = \bigotimes_{i = 1}^L\left|\downarrow\right\rangle, \qquad S^+\left|\downarrow\right\rangle = \left|\uparrow\right\rangle \quad \& \quad S^-\left|\uparrow\right\rangle = \left|\downarrow\right\rangle,
\end{IEEEeqnarray}
while the matrix $\theta_{jk}$ and the normalization constant $\mathfrak{N}$ are given by:
\begin{IEEEeqnarray}{c}
e^{i\theta_{jk}} \equiv \frac{u_j - u_k + i}{u_j - u_k - i} = S_{jk}, \qquad \mathfrak{N} \equiv \exp\left[-\frac{i}{2}\sum_{j<k}\theta_{jk}\right],
\end{IEEEeqnarray}
where we have omitted the level subscripts from the Bethe roots. $\mathfrak{N}$ is not present in \eqref{NestedWavefunction1}--\eqref{NestedWavefunction2}; it has been chosen in such a way that the coefficient of the term $e^{i p_k x_k}$ in \eqref{SU2wavefunction} is equal to one.
\paragraph{The $SU\left(3\right)$ sector} The $SU\left(3\right)$ subsector contains the three holomorphic fields $Z$, $W$ and $Y$. Once more the trace operator $\mathbb{K}$ does not contribute and the mixing matrix is the Heisenberg Hamiltonian \eqref{HeisenbergXXXonehalf}. The wavefunction $\psi\left(\textbf{u}_1,\textbf{u}_2\right)$ in \eqref{NestedBetheEigenstates} takes the following form:
\begin{IEEEeqnarray}{l}
\psi\left(\textbf{u},\textbf{v}\right) = \sum_{\sigma_1, \sigma_2} A_1\left(\sigma_1\right) \cdot A_2\left(\sigma_2\right) \cdot \prod_{j=1}^{M} \prod_{k=1}^{N_+} \left(\frac{u_{\sigma_{1,j}} + \frac{i}{2}}{u_{\sigma_{1,j}} - \frac{i}{2}}\right)^{n_{1,j}} \times \prod_{r=1}^{n_{2,k}} \frac{\left(v_{\sigma_{2,k}} - u_{\sigma_{1,r}} + \frac{i}{2}\right)^{\delta_{r\neq n_{2,k}}}}{v_{\sigma_{2,k}} - u_{\sigma_{1,r}} - \frac{i}{2}}, \qquad \label{SU3wavefunction}
\end{IEEEeqnarray}
where $M = N_1$ and $N_+ = N_2$ give the total number of level-1 and level-2 excitations of the spin chain. In \eqref{SU3wavefunction} we have also set $\textbf{u}_1 = \textbf{u} = \left\{u_1,\ldots,u_M\right\}$ and $\textbf{u}_2 = \textbf{v} = \left\{v_1,\ldots,v_{N_+}\right\}$.
\section[Solving the D3-D5 defect]{Solving the D3-D5 defect}
\noindent All the necessary tools are now available for the computation of the tree-level one-point function structure constants \eqref{OnePointFunctionsDefectCFTs}, \eqref{OnePointFunctionStructureConstant} for highest-weight eigenstates (Bethe states) of the $\mathfrak{so}\left(6\right)$ spin chain \eqref{DilatationOperator} and the matrix product states \eqref{MatrixProductStates}, in the planar limit ($N \rightarrow \infty$). For the dCFT that is dual to the D3-probe-D5 brane system, the form of the corresponding interface \eqref{FuzzyFunnelSU2} implies that the three holomorphic fields ($X, \ Y, \ Z$) are identical to the anti-holomorphic ones ($\bar{X}, \ \bar{Y}, \ \bar{Z}$).
\subsection[One-point functions in the $SU\left(2\right)$ sector]{One-point functions in the $SU\left(2\right)$ sector}
\noindent The overlap of the D3-D5 matrix product state in \eqref{MatrixProductStates} with an $M$-magnon Bethe state \eqref{SU2wavefunction} is the basic ingredient of the one-point function structure constant \eqref{OnePointFunctionStructureConstant}. It is given by
\begin{IEEEeqnarray}{c}
\left\langle\text{MPS}|\textbf{p}\right\rangle = \mathfrak{N} \cdot \sum_{\sigma \in S_M}\sum_{1\leq x_k \leq L} \exp\left[i\sum_{k} p_{\sigma_k}x_k + \frac{i}{2} \sum_{j<k} \theta_{\sigma_j\sigma_k}\right]\cdot\text{tr}\left[t_3^{x_1-1}t_1t_3^{x_2-x_1-1}\ldots\right],
\end{IEEEeqnarray}
where $\sigma$ is a permutation of $M$ letters and $p_j$ are the magnon momenta.
\paragraph{Selection rules} There exists a number of nifty selection rules for one-point functions in the $SU(2)$ sector \cite{deLeeuwKristjansenZarembo15}. In particular, $C_k$ has been found to vanish for all values of the bond dimension $k$ unless:
\begin{itemize}
\item The length of the operator $L$ and the number of magnons/excitations $M$ is even.
\item The magnon total momentum is zero, $\sum_{j = 1}^{M} p_j = 0$.
\item The Bethe roots $u_j$ are fully balanced: $\left\{u_1,\ldots,u_{M/2},-u_1,\ldots,-u_{M/2}\right\}$.
\end{itemize}
The first of the above properties is proven by means of certain similarity transformations that furnish an automorphism of the $\mathfrak{su}(2)$ algebra, while the second follows from trace cyclicity. The third selection rule is a consequence of the fact that the third conserved charge $\mathbb{Q}_3 = \sum_{j=1}^L \left[\mathbb{H}_{j-1,j},\mathbb{H}_{j,j+1}\right]$ in the integrable hierarchy of conserved mutually commuting charges of the spin chain \eqref{HeisenbergXXXonehalf} annihilates the matrix product state, i.e.\ $\mathbb{Q}_3\cdot\left|\text{MPS}\right\rangle = 0$. More about this property will be said in \S\ref{Section:IntegrableDefects} below. Fully balanced Bethe roots always appear in equal and opposite pairs.
\paragraph{Vacuum overlap} The vacuum overlap and one-point function ($M = 0$) has the following form for any value of the bond dimension $k$ \cite{deLeeuwKristjansenZarembo15}:
\begin{IEEEeqnarray}{c}
\left\langle\text{MPS}|0\right\rangle = \text{tr}\left[t_3^L\right] = \zeta\left(-L,\frac{1-k}{2}\right) - \zeta\left(-L,\frac{1+k}{2}\right), \qquad \zeta\left(s,\text{a}\right) \equiv \sum_{n=0}^{\infty} \frac{1}{\left(n+\text{a}\right)^s}, \label{VacuumOverlapSU2}
\end{IEEEeqnarray}
where $\zeta\left(s,\text{a}\right)$ is the Hurwitz zeta function. The vacuum overlap is nonzero only when the length of the operator $L$ is even, in full accordance with the above selection rules.
\paragraph{Determinant formula} For $M$ fully balanced excitations the one-point function structure constant \eqref{OnePointFunctionStructureConstant} becomes \cite{deLeeuwKristjansenZarembo15, Buhl-MortensenLeeuwKristjansenZarembo15}:
\begin{IEEEeqnarray}{l}
C_k\left(u\right) \equiv \frac{\left\langle\text{MPS}|u\right\rangle_k}{\sqrt{\left\langle u|u\right\rangle}} = C_2\left(u\right) \times \sum_{j = (1-k)/2}^{(k-1)/2}j^L \cdot \left[\prod_{l=1}^{M/2}\frac{u_l^2\left(u_l^2 + \frac{k^2}{4}\right)}{\left[u_l^2 + \left(j - \frac{1}{2}\right)^2\right] \cdot \left[u_l^2 + \left(j + \frac{1}{2}\right)^2\right]}\right], \qquad \label{OnePointFunctionsD3D5su2}
\end{IEEEeqnarray}
where $C_2\left(u\right)$ is the two-magnon one-point function (modulo the overall factor $L^{-1/2}\left(8\pi^2/\lambda\right)^{L/2}$)
\begin{IEEEeqnarray}{l}
C_2\left(u\right) \equiv \frac{\left\langle\text{MPS}|u\right\rangle_{k=2}}{\sqrt{\left\langle u|u\right\rangle}} = \sqrt{\prod_{j=1}^{M/2}\frac{u_j^2 + \frac{1}{4}}{u_j^2}\cdot\frac{\det G^+}{\det G^-}}
\end{IEEEeqnarray}
and $u = \left\{u_1,\ldots,u_{M/2}\right\}$. The $M/2\times M/2$ matrices $G_{jk}^{\pm}$ and $K_{jk}^{\pm}$ are defined as
\begin{IEEEeqnarray}{l}
G_{jk}^{\pm} = \left(\frac{L}{u_j^2 + \frac{1}{4}} - \sum_n K_{jn}^+\right)\delta_{jk} + K_{jk}^{\pm} \qquad \& \qquad K_{jk}^{\pm} = \frac{2}{1 + \left(u_j - u_k\right)^2} \pm \frac{2}{1 + \left(u_j + u_k\right)^2}. \qquad \label{NormMatrices1}
\end{IEEEeqnarray}
\subsection[One-point functions in the $SU\left(3\right)$ sector]{One-point functions in the $SU\left(3\right)$ sector}
\noindent As we have explained, the $SU\left(3\right)$ sector contains the three holomorphic fields \eqref{HolomorphicFields} of the theory. The Bethe eigenstates are constructed with the nested Bethe ansatz \eqref{NestedBetheEigenstates}, \eqref{SU3wavefunction} so that there are two types of excitations ($W$, $Y$) and two levels of Bethe roots, $\textbf{u}$ and $\textbf{v}$. Let us first define the Baxter functions $Q$ and $R$ :
\begin{IEEEeqnarray}{c}
Q_a\left(x\right) = \prod_{i=1}^{N_a}\left(i\,x - u_{a,i}\right), \qquad R_a\left(x\right) = \prod_{i=1}^{2\lfloor N_a/2\rfloor} \left(i\,x - u_{a,i}\right), \qquad a = 1,2,3. \label{BaxterFunctions}
\end{IEEEeqnarray}
The one-point functions in the $SU\left(3\right)$ sector (modulo $L^{-1/2}\left(8\pi^2/\lambda\right)^{L/2}$) are given by \cite{deLeeuwKristjansenLinardopoulos18a}:
\begin{IEEEeqnarray}{c}
C_k\left(u;v\right) = \mathbb{T}_{k-1} \times Q_1\left(k/2\right) \times \sqrt{\frac{Q_1\left(0\right)Q_1\left(1/2\right)}{R_2\left(0\right)R_2\left(1/2\right)}\cdot\frac{\det G^+}{\det G^-}}. \label{OnePointFunctionsD3D5su3}
\end{IEEEeqnarray}
As before these are only nonzero when the Bethe roots are fully balanced; also, $u = \left\{u_1,\ldots,u_{M/2}\right\}$ and $v = \left\{v_1,\ldots,v_{\lfloor N_+/2\rfloor}\right\}$. The full set of selection rules can be found below. For the definitions of the matrices $G^{\pm}$ see \eqref{NormMatrices2}--\eqref{NormMatrices3}. Also,
\begin{IEEEeqnarray}{c}
\mathbb{T}_s = \sum_{q = -s/2}^{s/2} q^L \cdot \frac{Q_2\left(q\right)}{Q_1\left(q + \frac{1}{2}\right) Q_1\left(q - \frac{1}{2}\right)}.
\end{IEEEeqnarray}
The validity of \eqref{OnePointFunctionsD3D5su3} has been checked numerically for a plethora of $SU\left(3\right)$ states that were obtained from the cNBA \eqref{NestedBetheEigenstates}, \eqref{SU3wavefunction}. For $k = 2$, \eqref{OnePointFunctionsD3D5su3} reduces to the determinant formula of \cite{deLeeuwKristjansenMori16}. In the absence of $Y$ excitations ($N_+ = 0$), the $SU(2)$ determinant formula \eqref{OnePointFunctionsD3D5su2} is retrieved. A recent attempt to prove \eqref{OnePointFunctionsD3D5su3} using the representation theory of twisted Yangians can be found in \cite{deLeeuwGomborKristjansenLinardopoulosPozsgay19}.
\subsection[One-point functions in the $SO\left(6\right)$ sector]{One-point functions in the $SO\left(6\right)$ sector}
\noindent The $SO(6)$ sector contains all the holomorphic and anti-holomorphic fields \eqref{HolomorphicFields}-\eqref{AntiHolomorphicFields}. There are three types of excitations and three sets of Bethe roots which must be fully balanced for the one-point function to be nonzero. Writing $w_j \equiv u_{3,j}$ ($j = 1,\ldots,N_3 = N_-$) and $w = \left\{w_1,\ldots,w_{\lfloor N_-/2\rfloor}\right\}$ the 1-point functions in the $SO\left(6\right)$ sector are given, for all values of the bond dimension $k$, by \cite{deLeeuwKristjansenLinardopoulos18a}:
\begin{IEEEeqnarray}{c}
C_k\left(u;v;w\right) =
\mathbb{T}_{k-1} \times Q_1\left(k/2\right) \times \sqrt{\frac{Q_1\left(0\right) Q_1\left(1/2\right)}{R_2\left(0\right) R_2\left(1/2\right) R_3\left(0\right) R_3\left(1/2\right)} \cdot \frac{\det G^+}{\det G^-}}, \label{OnePointFunctionsD3D5so6}
\end{IEEEeqnarray}
where again we have omitted the overall factor $L^{-1/2}\left(8\pi^2/\lambda\right)^{L/2}$. The definition of the Baxter functions $Q$ and $R$ has been given in \eqref{BaxterFunctions}. The matrices $G_{jk}^{\pm}$ and $K_{jk}^{\pm}$ are defined as \cite{KristjansenMullerZarembo20}:\footnote{There is a subtlety in the following formula, namely $K^{+}_{ab,jk}$ must be multiplied with $1/2$ whenever $u_{b,k} = 0$.}
\begin{IEEEeqnarray}{ll}
G^{\pm}_{ab,jk} = \delta_{ab}\delta_{jk}\left[\frac{L \, q_a^2}{u_{a,j}^2 + \frac{q_a^2}{4}} - \sum_{c = 1}^{3}\sum_{l = 1}^{\lceil N_c/2 \rceil} K^{+}_{ac,jl}\right] + K^{\pm}_{ab,jk}, \qquad &K^{\pm}_{ab,jk} = \mathbb{K}^{-}_{ab,jk} \pm \mathbb{K}^{+}_{ab,jk} \label{NormMatrices2} \\
& \mathbb{K}^{\pm}_{ab,jk} \equiv \frac{M_{ab}}{\left(u_{a,j} \pm u_{b,k}\right)^2 + \frac{1}{4}M_{ab}^2}, \qquad \label{NormMatrices3}
\end{IEEEeqnarray}
where $M_{ab}$ is the Cartan matrix and $q_a$ the highest-weight of $\mathfrak{so}\left(6\right)$ defined in \eqref{CartanMatrixSO6}. The indices $j,k$ of $K^{+}_{ab,jk}$ in \eqref{NormMatrices2}--\eqref{NormMatrices3} run from 1 to $\lceil N_{b}\rceil$ while those of $K^{-}_{ab,jk}$ run from 1 to $\lfloor N_{b}\rfloor$. \eqref{NormMatrices2}--\eqref{NormMatrices3} equally apply to the $\mathfrak{su}\left(2\right)$ and $\mathfrak{su}\left(3\right)$ algebras, reducing to \eqref{NormMatrices1} in the former case. An equivalent definition of the norm matrices $G_{jk}^{\pm}$ is given in appendix \ref{Appendix:BetheStateNorm}. The prefactor $\mathbb{T}$ in \eqref{OnePointFunctionsD3D5so6} can be related to the transfer matrix of the $\mathfrak{so}\left(6\right) \simeq \mathfrak{su}\left(4\right)$ spin chain. More will be said in \S\ref{FusionHierarchy} below. Explicitly,
\begin{IEEEeqnarray}{c}
\mathbb{T}_s = \sum_{q = -s/2}^{s/2} q^L \cdot \frac{Q_2\left(q\right) Q_3\left(q\right)}{Q_1\left(q+\frac{1}{2}\right) Q_1\left(q-\frac{1}{2}\right)}. \label{TransferMatrixD3D5so6}
\end{IEEEeqnarray}
The determinant formula \eqref{OnePointFunctionsD3D5so6} has been verified numerically for a large number of $SO(6)$ states that were constructed from the cNBA \eqref{NestedBetheEigenstates}, \eqref{NestedWavefunction1}--\eqref{NestedWavefunction2}. The $SU(2)$ and $SU(3)$ formulas \eqref{OnePointFunctionsD3D5su2} and \eqref{OnePointFunctionsD3D5su3} are trivially recovered when the corresponding excitations are absent ($N_{\pm} = 0$ or $N_{-} = 0$).
\paragraph{Selection rules} Once more, the one-point functions in the $SO(6)$ sector have been found to vanish for all values of the bond dimension $k$ unless:
\begin{itemize}
\item The quantities $M$ or $L + N_+ + N_-$ are both even.
\item The magnon total momentum is zero, $\sum_{j = 1}^{M} p_j = 0$.
\item All the levels of Bethe roots are fully balanced, i.e.\ they show up in equal and opposite pairs when the corresponding number of roots ($M$, $N_{\pm}$) is even and the extra unpaired root vanishes, when the corresponding wing root number ($N_{\pm}$) is odd:
\begin{IEEEeqnarray}{ll}
\left\{u_1,\ldots,u_{M/2},-u_1,\ldots,-u_{M/2}\right\}, \qquad & \left\{v_1,\ldots,v_{\lfloor N_+/2 \rfloor},-v_1,\ldots,-v_{\lfloor N_+/2\rfloor},0\right\} \label{BalancedBetheRoots1} \\[6pt]
& \left\{w_1,\ldots,w_{\lfloor N_-/2 \rfloor},-w_1,\ldots,-w_{\lfloor N_-/2\rfloor},0\right\}. \qquad \label{BalancedBetheRoots2}
\end{IEEEeqnarray}
\end{itemize}
The selection rules trivially reduce to the $SU(2)$ ones when $N_{\pm} = 0$. They are also fully valid in the $SU(3)$ sector, i.e.\ when $N_{-} = 0$.
\section[Solving the D3-D7 defect]{Solving the D3-D7 defect}
\noindent The dCFT that is dual to the D3-probe-D7 brane system is an example of a non-supersymmetric defect for which a determinant formula for all its scalar tree-level one-point functions of Bethe states has been found. This dCFT is drastically different from the supersymmetric D3-D5 dCFT that we examined in the previous section. Not only the closed-form expression for its one-point functions is more involved, but also the corresponding selection rules and the overall structure of its scalar subsectors are different. \\[6pt]
\indent The D3-D7 interface is described by the MPS in \eqref{MatrixProductStates} which is parameterized in terms of the $G$-matrices. The corresponding bond dimension $k = d_G$ in \eqref{InstantonNumber} is a function of the parameter $n$. As before, we also set $M = N_1$, $N_+ = N_2$ and $N_- = N_3$ for the $\mathfrak{so}\left(6\right)$ excitation mode numbers.
\paragraph{Vacuum overlap} For the vacuum overlap ($M=0$) we have found \cite{deLeeuwKristjansenLinardopoulos16}:
\begin{IEEEeqnarray}{ll}
\left\langle\text{MPS}|0\right\rangle = \text{Tr}\left[G_5^L\right] = 2^L \times \Bigg\{\frac{\left(n+2\right)^2}{4} \cdot \Big[&\zeta\left(-L,-\frac{n}{2}\right) - \zeta\left(-L,\frac{n}{2}+1\right)\Big] - \nonumber \\
& - \Big[\zeta\left(-L-2,-\frac{n}{2}\right) - \zeta\left(-L-2,\frac{n}{2}+1\right)\Big]\Bigg\}, \qquad
\end{IEEEeqnarray}
where the Hurwitz zeta function $\zeta\left(s,\text{a}\right)$ was defined in \eqref{VacuumOverlapSU2}. By using the relationship between $\zeta\left(s,\text{a}\right)$ and the Bernoulli polynomials $B_{m}\left(x\right)$ the vacuum overlap can also be written as
\begin{IEEEeqnarray}{c}
\left\langle\text{MPS}|0\right\rangle = \left\{\begin{array}{l} 0, \quad L \ \text{odd} \\[6pt] 2^L \times \left[\frac{2}{L+3} \, B_{L+3}\left(-\frac{n}{2}\right) - \frac{\left(n+2\right)^2}{2\left(L+1\right)} \, B_{L+1}\left(-\frac{n}{2}\right)\right], \quad L \ \text{even}. \end{array}\right.
\end{IEEEeqnarray}
\paragraph{Selection rules} As it turns out, the overlaps $\left\langle\text{MPS}|\Psi\right\rangle$ of all the highest-weight eigenstates $\left|\Psi\right\rangle$ vanish unless the number of holomorphic excitations $W$ and $Y$ in $\left|\Psi\right\rangle$ is respectively equal to the number of the anti-holomorphic fields $\bar{W}$ and $\bar{Y}$. \eqref{HolomporphicFieldExcitations}--\eqref{AntiholomporphicFieldExcitations} then directly imply that one-point functions vanish for all values of the parameter $n$ in the $SU\left(2\right)$ and $SU\left(3\right)$ subsectors and only the vacuum state and the $SO(6)$ sector can be nontrivial. \\[6pt]
\indent Moreover, the $SO(6)$ one-point functions have been found to vanish for all the values of the parameter $n$ unless:
\begin{itemize}
\item The excitation mode numbers satisfy $M = 2N_+ = 2N_-$ and $M$ is even.
\item The magnon total momentum is zero, $\sum_{j = 1}^{M} p_j = 0$.
\item All the levels of Bethe roots are fully balanced as in \eqref{BalancedBetheRoots1}--\eqref{BalancedBetheRoots2}.
\item $L$ is odd and the wing rapidities are not identical, $v \neq w$.
\end{itemize}
\paragraph{Determinant formula}
Using the Baxter functions $Q$ and $R$ that have been defined in \eqref{BaxterFunctions}, setting $u_i \equiv u_{1,i}, \ v_j \equiv u_{2,j}, \ w_l \equiv u_{3,l}$ and also $u = \left\{u_1,\ldots,u_{M/2}\right\}$, $v = \left\{v_1,\ldots,v_{\lfloor N_+/2\rfloor}\right\}$, $w = \left\{w_1,\ldots,w_{\lfloor N_-/2\rfloor}\right\}$ for the balanced Bethe roots, we find the following expression for one-point functions in the $SO(6)$ sector \cite{deLeeuwGomborKristjansenLinardopoulosPozsgay19}:
\begin{IEEEeqnarray}{c}
C_n\left(u;v;w\right) =
\mathbb{T}_{n} \cdot \sqrt{\frac{Q_1\left(0\right) Q_1\left(1/2\right)}{R_2\left(0\right) R_2\left(1/2\right) R_3\left(0\right) R_3\left(1/2\right)} \cdot \frac{\det G^+}{\det G^-}}, \label{OnePointFunctionsD3D7so6}
\end{IEEEeqnarray}
omitting as usual the factor $L^{-1/2}\left(8\pi^2/\lambda\right)^{L/2}$. The definition of the matrices $G^{\pm}$ has been given above in \eqref{NormMatrices2}--\eqref{NormMatrices3}. See also \eqref{NormMatrices4}. The prefactor $\mathbb{T}$ is given by
\begin{IEEEeqnarray}{c}
\mathbb{T}_n = \sum_{q = -n/2}^{n/2} \left(2q\right)^L
\Bigg[\sum_{p=-n/2}^q \frac{Q_{1}\left(p-\frac{1}{2}\right)}{Q_{1}\left(q-\frac{1}{2}\right)}\frac{Q_{3}\left(q\right) Q_{3}\left(\frac{n}{2}+1\right)}{Q_{3}\left(p\right) Q_{3}\left(p-1\right)}\Bigg]
\Bigg[\sum_{r=q}^{n/2} \frac{Q_{1}\left(r+\frac{1}{2}\right)}{Q_{1}\left(q+\frac{1}{2}\right)}\frac{Q_{2}\left(q\right) Q_{2}\left(\frac{n}{2}+1\right)}{Q_{2}\left(r\right) Q_{2}\left(r+1\right)}\Bigg] \qquad \label{TransferMatrixD3D7even}
\end{IEEEeqnarray}
however this expression is valid only for even values of $(n + 1)M/2$. When $(n + 1)M/2$ is odd, one auxiliary Bethe root vanishes at each level and a $0/0$ indeterminate form arises. In this case $\mathbb{T}_n$ becomes
\begin{IEEEeqnarray}{ll}
\mathbb{T}_n = &\sum_{q = 0}^{n/2} \left(2q\right)^L \, \Lambda_n^+(q) \Bigg[\Lambda_n^-(q) + \frac{Q_1\left(1/2\right) Q_3(q) Q_3\left(\frac{n}{2} + 1\right)}{Q_1\left(q - \frac{1}{2}\right) Q_3(1) R_3(0)} \cdot \frac{d}{du}\ln\left[\frac{Q_1\left(1/2\right)}{Q_3(1)}\right]^2\Bigg] + \nonumber \\
& + \sum_{q = -n/2}^{0} \left(2q\right)^L \, \Lambda_n^-(q) \Bigg[\Lambda_n^+(q) + \frac{Q_1\left(1/2\right) Q_2(q) Q_2\left(\frac{n}{2} + 1\right)}{Q_1\left(q + \frac{1}{2}\right) Q_2(1) R_2(0)} \cdot \frac{d}{du}\ln\left[\frac{Q_1\left(1/2\right)}{Q_2(1)}\right]^2\Bigg], \qquad \label{TransferMatrixD3D7odd1}
\end{IEEEeqnarray}
where $\Lambda_n^{\pm}(q)$ are given by
\begin{IEEEeqnarray}{c}
\Lambda_n^+(q) \equiv \sum_{\substack{r = q \\ r \neq 0, -1}}^{n/2} \frac{Q_1\left(r + \frac{1}{2}\right)Q_2(q) Q_2\left(\frac{n}{2} + 1\right)}{Q_1\left(q + \frac{1}{2}\right)Q_2(r + 1) Q_2(r)}, \
\Lambda_n^-(q) \equiv \sum_{\substack{r = -n/2 \\ r \neq 0, 1}}^{q} \frac{Q_1\left(r - \frac{1}{2}\right) Q_3(q) Q_3\left(\frac{n}{2} + 1\right)}{Q_1\left(q - \frac{1}{2}\right) Q_3(r) Q_3(r - 1)}. \qquad \label{TransferMatrixD3D7odd2}
\end{IEEEeqnarray}
The determinant formulas \eqref{OnePointFunctionsD3D7so6}--\eqref{TransferMatrixD3D7odd2} have been checked numerically for many $SO(6)$ states constructed by means of the nested Bethe ansatz \eqref{NestedBetheEigenstates}, \eqref{NestedWavefunction1}--\eqref{NestedWavefunction2}. For $(M,N_{+},N_{-}) = (2,1,1)$ \eqref{OnePointFunctionsD3D7so6}--\eqref{TransferMatrixD3D7odd2} trivially reduce to the corresponding formula of \cite{deLeeuwKristjansenLinardopoulos16}. \\[6pt]
\indent Here are some interesting special cases in which the formulae \eqref{TransferMatrixD3D7even}--\eqref{TransferMatrixD3D7odd2} simplify. For $n = 1$, \eqref{TransferMatrixD3D7even} reads
\begin{IEEEeqnarray}{c}
\mathbb{T}_1=\left(1 + (-1)^L\right) \cdot \frac{Q_1\left(1\right)}{Q_1\left(0\right)} + (-1)^{N_-} \cdot \frac{Q_3\left(3/2\right)}{Q_3\left(1/2\right)} + (-1)^{L + N_+} \cdot \frac{Q_2\left(3/2\right)}{Q_2\left(1/2\right)},
\end{IEEEeqnarray}
whereas for $n = 2$, \eqref{TransferMatrixD3D7even}--\eqref{TransferMatrixD3D7odd2} take the following form
\begin{IEEEeqnarray}{ll}
\mathbb{T}_2 = 2^{L+1} \times \Bigg\{\frac{\left(1 + (-1)^L\right)}{2} \cdot \frac{Q_1\left(3/2\right)}{Q_1\left(1/2\right)} &+ \frac{Q_3(2)}{R_3(0)}\left[\frac{Q'_1\left(1/2\right)}{Q_1\left(1/2\right)} - \frac{Q'_3(1)}{Q_3(1)}\right]^{\delta_{M/2 = \text{odd}}} + \nonumber \\
& + (-1)^{L} \cdot \frac{Q_2(2)}{R_2(0)}\left[\frac{Q'_1\left(1/2\right)}{Q_1\left(1/2\right)} - \frac{Q'_2(1)}{Q_2(1)}\right]^{\delta_{M/2 = \text{odd}}}\Bigg\}. \qquad
\end{IEEEeqnarray}
The prefactors $\mathbb{T}$ that show up in the one-point function determinant formulas \eqref{OnePointFunctionsD3D5so6}, \eqref{OnePointFunctionsD3D7so6} can again be related to the eigenvalues of the transfer matrix of the $\mathfrak{so}\left(6\right) \simeq \mathfrak{su}\left(4\right)$ spin chain. We briefly explore this connection in the next section.
\section[Fusion hierarchies]{Fusion hierarchies \label{FusionHierarchy}}
\noindent Every irreducible representation of $GL\left(N\right)$ that is parameterized by a semistandard Young tableau\footnote{Semistandard means that the entries of the YT increase weakly along each row and strictly along each column.} with $a$ rows and $s$ columns gives rise to a transfer matrix (TM) of the $\mathfrak{gl}\left(N\right)$ spin chain. It can be shown that the eigenvalue $T_{a,s}$ of this TM satisfies the Hirota equation or T-system:
\begin{IEEEeqnarray}{c}
T_{a,s}\left(u + \frac{i}{2}\right) T_{a,s}\left(u - \frac{i}{2}\right) = T_{a+1,s}\left(u\right) T_{a-1,s}\left(u\right) + T_{a,s+1}\left(u\right) T_{a,s-1}\left(u\right), \label{T-system}
\end{IEEEeqnarray}
a difference equation that describes the fusion of transfer matrices ($u$ is known as the spectral parameter). An equivalent representation of \eqref{T-system} can be obtained by the transformation
\begin{IEEEeqnarray}{c}
Y_{a,s} = \frac{T_{a,s+1} T_{a,s-1}}{T_{a+1,s} T_{a-1,s}},
\end{IEEEeqnarray}
which leads to the so-called Y-system:
\begin{IEEEeqnarray}{c}
Y_{a,s}^+ Y_{a,s}^- = \frac{\left(1 + Y_{a,s+1}\right) \left(1 + Y_{a,s-1}\right)}{\left(1 + 1/Y_{a+1,s}\right) \left(1 + 1/Y_{a-1,s}\right)}, \qquad Y_{a,s}^{\pm} \equiv Y_{a,s}\left(u \pm \frac{i}{2}\right). \label{Y-system}
\end{IEEEeqnarray}
Nothing much changes in the above for $\mathfrak{su}\left(N\right)$ spin chains obeying periodic or open boundary conditions (and the TM is single-row or double-row respectively). The T-system \eqref{T-system} affords a solution that can be expressed by the following tableau sum formula \cite{KunibaNakanishiSuzuki10}
\begin{IEEEeqnarray}{c}
T_{a,s}\left(i \, u\right) = \sum_{\varpi} \ \mathop{\prod_{l = 1,\ldots, a}}_{m = 1,\ldots, s} z_{(\varpi_{lm})} \left[u + \frac{1}{2} \, \left(a - 2l - s + 2m\right)\right], \label{TableauSumFormula}
\end{IEEEeqnarray}
where the indices $a = 1,\ldots, N-1$ and $s = 1,2,\ldots$ correspond to the rows and the columns of the semistandard Young tableau that parameterize the irreducible representation of $SU(N)$. The prefactors $\mathbb{T}$ in \eqref{OnePointFunctionsD3D5so6}, \eqref{OnePointFunctionsD3D7so6} are related to the $a = 1$ transfer matrices. For $a = 1$ the tableau sum formula \eqref{TableauSumFormula} becomes
\begin{IEEEeqnarray}{c}
T_{1,s}\left(i \, u\right) = \sum_{\varpi} \prod_{m = 1}^{s} z_{(\varpi_{m})}\left[u - \frac{1}{2} \, \left(s - 2m + 1\right)\right], \label{TableauSumFormula1}
\end{IEEEeqnarray}
where the sum runs over all the $\mathfrak{su}\left(N\right)$ semistandard Young tableaux $\varpi$ of size $1\times s$.
\subsection[D3-D5 system]{D3-D5 system}
\noindent Let us see how the $a = 1$ transfer matrix eigenvalue \eqref{TableauSumFormula1} is related to the prefactor \eqref{TransferMatrixD3D5so6}. Plugging the following set of $z$-functions
\begin{IEEEeqnarray}{ll}
z_{1}(u) & = i\,\left(u + \frac{s - 1}{2}\right) \cdot \frac{Q_1\left(u - 1\right)}{Q_1\left(u\right)} \label{HirotaD3D5-1} \\[6pt]
z_{2}(u) & = \frac{Q_1\left(u - 1\right)}{Q_1\left(u\right)} \label{HirotaD3D5-2} \\[6pt]
z_{3}(u) & = \frac{\left(u - \frac{1}{2}\right)^L}{\left(u + \frac{1}{2}\right)^L} \cdot \frac{Q_1\left(u + 1\right)}{Q_1\left(u\right)} \cdot \frac{Q_2\left(u - \frac{1}{2}\right) Q_3\left(u - \frac{1}{2}\right)}{Q_2\left(u + \frac{1}{2}\right) Q_3\left(u + \frac{1}{2}\right)} \label{HirotaD3D5-3} \\[6pt]
z_{4}(u) & = i\,\left(u - \frac{s - 1}{2}\right) \cdot \frac{\left(u - \frac{1}{2}\right)^L}{\left(u + \frac{1}{2}\right)^L} \cdot \frac{Q_1\left(u + 1\right)}{Q_1\left(u\right)} \cdot \frac{Q_2\left(u - \frac{1}{2}\right) Q_3\left(u - \frac{1}{2}\right)}{Q_2\left(u + \frac{1}{2}\right) Q_3\left(u + \frac{1}{2}\right)}, \label{HirotaD3D5-4}
\end{IEEEeqnarray}
into the $N = 4$ (i.e.\ $\mathfrak{su}\left(4\right)$) tableau sum formula \eqref{TableauSumFormula1}, we obtain:
\begin{IEEEeqnarray}{c}
T_{1,s}\left(0\right) = \frac{Q_1^2\left((s+1)/2\right)}{Q_2\left(s/2\right) Q_3\left(s/2\right)} \times \sum_{q = -s/2}^{s/2} \left(-\frac{2q}{s}\right)^L \cdot \frac{Q_2\left(q\right) Q_3\left(q\right)}{Q_1\left(q+\frac{1}{2}\right) Q_1\left(q-\frac{1}{2}\right)}.
\end{IEEEeqnarray}
Therefore the following relation between the prefactor \eqref{TransferMatrixD3D5so6} and the TM eigenvalue $T_{1,s}\left(0\right)$ holds:
\begin{IEEEeqnarray}{c}
\mathbb{T}_s\left[\text{D3-D5}\right] = \left(-\frac{s}{2}\right)^L \cdot \frac{Q_2\left(s/2\right) Q_3\left(s/2\right)}{Q_1^2\left((s+1)/2\right)} \times T_{1,s}\left(0\right).
\end{IEEEeqnarray}
\subsection[D3-D7 system]{D3-D7 system}
\noindent Similarly, by plugging the following set of $z$-functions \cite{deLeeuwGomborKristjansenLinardopoulosPozsgay19}
\begin{IEEEeqnarray}{ll}
z_{1}(u) & = \frac{(u + \frac{1}{2})^{L}}{(u - \frac{1}{2})^{L}} \cdot \frac{Q_{2}(u - \frac{3}{2})}{Q_{2}(u - \frac{1}{2})} \label{HirotaD3D7-1} \\[6pt]
z_{2}(u) & = \frac{(u + \frac{1}{2})^{L}}{(u - \frac{1}{2})^{L}} \cdot \frac{Q_{1}(u - 1)}{Q_{1}(u)} \cdot \frac{Q_{2}(u + \frac{1}{2})}{Q_{2}(u - \frac{1}{2})} \label{HirotaD3D7-2} \\[6pt]
z_{3}(u) & = \frac{Q_{1}(u + 1)}{Q_{1}(u)} \cdot \frac{Q_{3}(u - \frac{1}{2})}{Q_{3}(u + \frac{1}{2})} \label{HirotaD3D7-3} \\[6pt]
z_{4}(u) & = \frac{Q_{3}(u + \frac{3}{2})}{Q_{3}(u + \frac{1}{2})}, \label{HirotaD3D7-4}
\end{IEEEeqnarray}
into the $\mathfrak{su}\left(4\right)$ tableau sum formula \eqref{TableauSumFormula1} we are led to the following expression
\begin{IEEEeqnarray}{c}
T_{1,n} = \sum_{q = -\frac{n}{2}}^{\frac{n}{2}} \left(2q\right)^L
\Bigg[\sum_{p=-\frac{n}{2}}^q \frac{Q_{1}\left(p-\frac{1}{2}\right)}{Q_{1}\left(q-\frac{1}{2}\right)}\frac{Q_{3}\left(q\right) Q_{3}\left(\frac{n}{2}+1\right)}{Q_{3}\left(p\right) Q_{3}\left(p-1\right)}\Bigg]
\Bigg[\sum_{r=q}^{\frac{n}{2}} \frac{Q_{1}\left(r+\frac{1}{2}\right)}{Q_{1}\left(q+\frac{1}{2}\right)}\frac{Q_{2}\left(q\right) Q_{2}\left(\frac{n}{2}+1\right)}{Q_{2}\left(r\right) Q_{2}\left(r+1\right)}\Bigg]. \qquad
\end{IEEEeqnarray}
The prefactor \eqref{TransferMatrixD3D7even} is related to the TM eigenvalue $T_{1,n}\left(0\right)$ as
\begin{IEEEeqnarray}{c}
\mathbb{T}_n\left[\text{D3-D7}\right] = \left(-n\right)^L \times T_{1,n}\left(0\right).
\end{IEEEeqnarray}
\section[Integrable defects]{Integrable defects \label{Section:IntegrableDefects}}
\noindent There is an interesting argument that allows to view defects as quantum quenches and vice-versa. According to Ghoshal and Zamolodchikov \cite{GhoshalZamolodchikov93}, there are two equivalent ways to describe an (integrable) QFT in the presence of a boundary/defect, one in which the defect operates as a wall or boundary condition which scatters off particles, and another (in the Wick-rotated frame of the first) where the defect operates as an initial condition for the pair-creation/annihilation of particles (see figure \ref{Figure:GhoshalZamolodchikov} below).
\begin{figure}[H]\begin{center}
\includegraphics[scale=1]{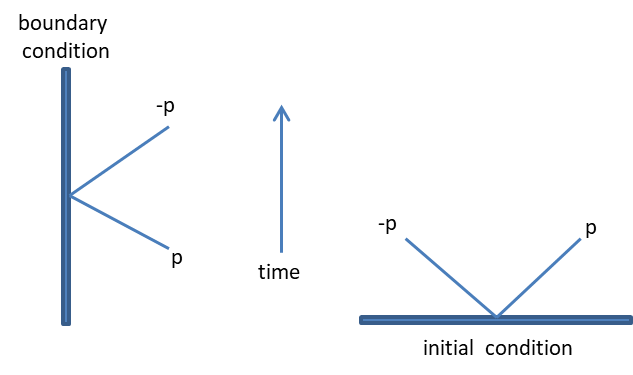}
\caption{Quenches and defects in the rotated channel.} \label{Figure:GhoshalZamolodchikov}
\end{center}\vspace{-1.5em}\end{figure}
\paragraph{Quantum quenches} To further illustrate this point, let us first explain the notion of (global) quantum quench. Consider an isolated quantum many-particle system that is described in terms of a translationally-invariant, time-independent Hamiltonian $\mathbb{H}(h)$ with short-range interactions ($h$ is a tunable parameter of the Hamiltonian, e.g.\ a coupling constant). At time $t < 0$ the system is prepared to the ground state $\left|B\left(0\right)\right\rangle$ of $\mathbb{H}(h_0)$. Then suddenly at $t = 0$ the parameter $h_0$ is "quenched" (i.e.\ changed) to a new value $h$ without affecting the initial state ("sudden approximation"). For $t > 0$ the system evolves unitarily according to the Schr\"{o}dinger equation
\begin{IEEEeqnarray}{c}
\left|B\left(t\right)\right\rangle = e^{-i \, \mathbb{H}(h) \, t}\left|B\left(0\right)\right\rangle \label{InitialStateEvolution}
\end{IEEEeqnarray}
and may or may not equilibrate to a stationary state as $t \rightarrow \infty$. The restriction to short-range interactions (e.g.\ spin chains) is important given that systems with infinitely-ranged interactions have a different equilibration behavior. On the other hand there exist more quench protocols, e.g.\ when the initial state is not pure, or when translation invariance is limited to only a few sites of the system. The study of quantum quenches has exploded in recent years with a series of fascinating experiments involving ultra-cold trapped atoms (see e.g.\ \cite{LangenGasenzerSchmiedmayer16} and references therein). \\[6pt]
\indent The evolution of the quenched system from small to large time scales is best studied through the time-dependent expectation values of various operators. These can often be related to overlaps of matrix product states and spin chain eigenstates like the ones we computed in the previous sections. To see this let us write the initial state $\left|B\left(0\right)\right\rangle$ in \eqref{InitialStateEvolution} as a matrix product state $\left|B\left(0\right)\right\rangle = \left|\text{MPS}\right\rangle$. The initial states are prepared as ground states of some Hamiltonian that is afterwards quenched to a periodic spin chain like \eqref{DilatationOperator}, \eqref{HeisenbergXXXonehalf}. The quenched Hamiltonian $\mathbb{H}$ can be diagonalized as
\begin{IEEEeqnarray}{c}
\mathbb{H}\left|\Psi_n\right\rangle = E_n\left|\Psi_n\right\rangle \label{QuenchedHamiltonianEigenstates}
\end{IEEEeqnarray}
while the evolution equation \eqref{InitialStateEvolution} takes the form
\begin{IEEEeqnarray}{c}
\left|B\left(t\right)\right\rangle = \sum_{n} e^{-i E_{n}t} \cdot \left\langle\Psi_n|\text{MPS}\right\rangle \cdot \left|\Psi_n\right\rangle.
\end{IEEEeqnarray}
The post-quench expectation values of generic physical operators $O$ then become
\begin{IEEEeqnarray}{c}
\left\langle B\left(t\right)\right| O \left|B\left(t\right)\right\rangle = \sum_{m,n} \left\langle\Psi_n|\text{MPS}\right\rangle \left\langle \text{MPS}| \Psi_m \right\rangle \left\langle \Psi_m \right| O \left|\Psi_n\right\rangle \cdot e^{-i \left(E_{n} - E_{m}\right) t},
\end{IEEEeqnarray}
where $\left\langle \text{MPS}| \Psi_n \right\rangle$ is the overlap of the initial state $\left|B\left(0\right)\right\rangle = \left|\text{MPS}\right\rangle$ with the eigenstates $\left|\Psi_n\right\rangle$ of the quenched Hamiltonian $\mathbb{H}$. \\[6pt]
\indent Now it is known that in the absence of other conserved quantities, isolated Hamiltonian systems always relax to a thermal equilibrium, i.e.\ they thermalize. The stationary state is described by the Gibbs ensemble. In the presence of more conserved quantities, systems are known to relax to a generalized Gibbs ensemble (GGE). Integrable models demonstrate an unusual equilibration dynamics: they do not thermalize.\footnote{More about this interesting topic can be found in the collection of articles \cite{CalabreseEsslerMussardo16}.} \\[6pt]
\indent Considerable effort has therefore been devoted to the study of relaxation dynamics of integrable models. In order to overcome a number of difficulties that arise when applying the GGE, \cite{PiroliPozsgayVernier17} proposed the study of quenches that arise from certain initial states that have been dubbed integrable initial states.
\paragraph{Note on integrability} Before proceeding to the definition of an integrable quench, let us briefly revisit the concept of integrability. Intuitively, integrability implies exact solvability. This follows from Liouville's theorem which states that the equations of motion of a Hamiltonian system having $M$ degrees of freedom and an equal number of Poisson-commuting conserved quantities $Q_a$
\begin{IEEEeqnarray}{c}
\left\{Q_a,Q_b\right\} = 0, \qquad a,b = 1,\ldots, M \label{ClassicalIntegrability}
\end{IEEEeqnarray}
is solvable by quadratures; $Q_1 = P$ is the total momentum, $Q_2 = H$ the Hamiltonian of the system and $\left\{*,*\right\}$ the Poisson bracket. In other words the solution can in principle be obtained by solving a finite number of algebraic equations and performing a finite number of integrations. Practically, there are very few integrable systems for which closed-form solutions can be found though. \\[6pt]
\indent The direct generalization of \eqref{ClassicalIntegrability} to quantum systems seems to be (1):
\begin{IEEEeqnarray}{c}
\left[\mathbb{Q}_a,\mathbb{Q}_b\right] = 0, \qquad a,b = 1,\ldots, M, \label{QuantumIntegrabilityFinite}
\end{IEEEeqnarray}
i.e.\ the classical charges are promoted to operators and Poisson brackets to commutators. This definition however turns out to be neither sufficiently rigorous nor practical. Other definitions such as (2) solvability by the Bethe ansatz (diagonalizability), (3) multi-particle scattering that is factorizable in two-particle scatterings (non-diffractive scattering), etc.\ are also not completely consistent as sufficient conditions, despite being for the most part accurate as necessary conditions for quantum integrability. To day there is no general consensus as to the proper rigorous necessary and sufficient conditions of quantum integrability.
\paragraph{Integrable quenches} To formulate the definition of an integrable quench we consider an integrable Hamiltonian system with infinitely many mutually commuting conserved local charges $\mathbb{Q}_a$
\begin{IEEEeqnarray}{c}
\left[\mathbb{Q}_a,\mathbb{Q}_b\right] = 0, \qquad a,b = 1,2,\ldots \label{QuantumIntegrabilityInfinite}
\end{IEEEeqnarray}
These charges can always be chosen to have a definite parity with respect to space reflections, i.e.\
\begin{IEEEeqnarray}{c}
\Pi \, \mathbb{Q}_a \, \Pi = \left(-1\right)^a \mathbb{Q}_a , \qquad a = 1,2\ldots, \label{SpaceParityOperator}
\end{IEEEeqnarray}
so that they can be divided into two classes according to their space parity, the even charges $\mathbb{Q}_{2n}$ and the odd charges $\mathbb{Q}_{2n+1}$. In \eqref{SpaceParityOperator}, $\Pi$ stands for the space parity operator
\begin{IEEEeqnarray}{c}
\Pi\left|i_1,i_2,\ldots,i_{L-1},i_L\right\rangle = \left|i_L,i_{L-1},\ldots,i_2,i_1\right\rangle.
\end{IEEEeqnarray}
An initial state $\left|\text{MPS}\right\rangle$ that is represented by a matrix product state is integrable if it is annihilated by all the parity-odd conserved charges of the integrable hierarchy \eqref{QuantumIntegrabilityInfinite} \cite{PiroliPozsgayVernier17}:
\begin{IEEEeqnarray}{c}
\mathbb{Q}_{2s+1}\left|\text{MPS}\right\rangle = 0, \qquad s = 1,2,\ldots \label{QuenchIntegrabilityCondition1}
\end{IEEEeqnarray}
In fact there is a stronger version of \eqref{QuenchIntegrabilityCondition1} that involves the transfer matrix $\tau$ of the system:
\begin{IEEEeqnarray}{c}
\Pi\,\tau\,\Pi\left|\text{MPS}\right\rangle = \tau\left|\text{MPS}\right\rangle. \label{QuenchIntegrabilityCondition2}
\end{IEEEeqnarray}
\indent The definitions \eqref{QuenchIntegrabilityCondition1}--\eqref{QuenchIntegrabilityCondition2} of integrable quenches have a number of immediate consequences for the form $\left\langle\text{MPS}|\Psi\right\rangle$ of matrix product state overlaps with the (Bethe) eigenstates of the Hamiltonian $\mathbb{H} = \mathbb{Q}_2$. For one, the overlaps vanish unless the rapidities of the excitations (Bethe roots) come in pairs of opposite signs as e.g.\ in \eqref{BalancedBetheRoots1}--\eqref{BalancedBetheRoots2}. Moreover, the overlaps can be described by closed-form expressions (as e.g.\ in \eqref{OnePointFunctionsD3D5so6}, \eqref{OnePointFunctionsD3D7so6}) with simple factorized descriptions.
\paragraph{Integrable defects} We can now go full circle by formulating the lattice version of the Ghoshal-Zamolodchikov rotation that was presented at the beginning of the section (see figure \ref{Figure:GhoshalZamolodchikov}). Following \cite{PiroliPozsgayVernier17} the initial states of closed spin chains are mapped to boundary states of open spin chains. The latter are determined in terms of a reflection matrix $K$ which provides the absorption amplitude of particle pairs scattering off the boundary. Integrable quenches satisfying \eqref{QuenchIntegrabilityCondition1}--\eqref{QuenchIntegrabilityCondition2} correspond to integrable boundaries whose reflection matrices $K$ satisfy the boundary Yang-Baxter (or reflection) equation \cite{PiroliPozsgayVernier17, PiroliVernierCalabresePozsgay18a, PiroliVernierCalabresePozsgay18b, PozsgayPiroliVernier18b}. \\[6pt]
\indent The integrable boundary state/reflection matrix controls the set of integrable boundary conditions that are obeyed by the open spin chains. Besides single-trace operators that are composed exclusively of bulk scalars $\phi$, defect CFTs with boundary degrees of freedom like \eqref{ActiondCFT}, also have operators that contain defect fields, e.g.\ $O = \hat{\phi}^{\dag}\phi\phi\ldots\phi\hat\phi$, where $\hat{\phi}$ is a defect scalar. The set of integrable spin chains with open boundary conditions that give the corresponding one-loop dilatation operator of the D3-D5 interface has been studied in \cite{DeWolfeMann04, IpsenVardinghus19}. \\[6pt]
Here is a summary of the main properties of integrable quenches/defects:
\begin{itemize}
\item The parity-odd conserved charges annihilate the MPS, \eqref{QuenchIntegrabilityCondition1}.
\item The transfer matrix obeys \eqref{QuenchIntegrabilityCondition2}.
\item Boundary states corresponding to open spin chain boundary conditions are integrable.
\item Bethe roots are fully balanced, i.e.\ they appear in pairs of opposite signs.
\item Closed-form expressions are possible for overlaps of Bethe eigenstates with MPSs.
\end{itemize}
It has been proven in \cite{deLeeuwKristjansenLinardopoulos18a} that the matrix product states \eqref{MatrixProductStates} that are relevant to the D3-D5 and the D3-D7 interfaces are integrable, namely they satisfy the criteria \eqref{QuenchIntegrabilityCondition1}--\eqref{QuenchIntegrabilityCondition2}. Both interfaces afford closed-form expressions for all the tree-level one-point functions of their bulk single-trace scalar operators that are highest-weight states of the corresponding dilatation operators, cf.\ \eqref{OnePointFunctionsD3D5so6}, \eqref{OnePointFunctionsD3D7so6}. The other main consequence of integrability, the pairing of Bethe roots \eqref{BalancedBetheRoots1}--\eqref{BalancedBetheRoots2}, has also been checked for a large number of nontrivial operators for both interfaces in \eqref{MatrixProductStates}.
\section[Summary and outlook]{Summary and outlook}
\noindent Following the recent works \cite{deLeeuwKristjansenLinardopoulos18a, deLeeuwGomborKristjansenLinardopoulosPozsgay19}, we have presented closed-form expressions \eqref{OnePointFunctionsD3D5so6}, \eqref{OnePointFunctionsD3D7so6} for all the tree-level one-point functions of the defect CFTs that are dual to the $SU(2)$ symmetric D3-probe-D5 brane system with $k$ units of magnetic flux and the $SO(5)$ symmetric D3-probe-D7 brane system with $d_G$ units of instanton flux. Working in the planar limit ($N \rightarrow \infty$) and at weak 't Hooft coupling ($\lambda \rightarrow 0$), we concentrated on bulk single-trace scalar operators that are highest-weight (Bethe) states of the integrable $\mathfrak{so}(6)$ spin chain \eqref{DilatationOperator}. \\[6pt]
\indent Checking the validity of the determinant formulas \eqref{OnePointFunctionsD3D5su3}, \eqref{OnePointFunctionsD3D5so6}, \eqref{OnePointFunctionsD3D7so6} and the corresponding selection rules involves solving the Bethe equations \eqref{BetheEquations1}--\eqref{BetheEquations3}. The Bethe roots are used both in numerically obtaining the one-point functions from the overlaps of the nested Bethe states \eqref{NestedBetheEigenstates}, \eqref{NestedWavefunction1}--\eqref{NestedWavefunction2} and in computing their values from the aforementioned determinant formulas. For all the checks that are mentioned in the present talk we have used an algorithm (\textit{fast Bethe solver}) that was developed in \cite{MarboeVolin14, MarboeVolin17, Marboe17}. This program quickly solves the Bethe equations by demanding the solutions of the QQ-relation to be polynomials. \\[6pt]
\indent In order to come closer to solving the defects (in the sense explained in \S\ref{Section:BoundaryConformalFieldTheories}), more input is needed such as the data from the pure CFT ($\mathcal{N} = 4$ SYM), the boundary and the bulk-boundary data. As we mentioned in \S\ref{Subsection:DefectConformalFieldTheories} these data are not all independent; the boundary CFT bootstrap program aims to determine their precise relationship. This is certainly not the end of the story as there are more sectors in the theory and far more operators such as descendants and multiple-trace operators. Furthermore, integrability may be used to go beyond the tree-level approximation and the large-$N$ limit, as well as computing the values of more and more interesting observables. \\[6pt]
\indent Naturally, there are a lot of topics in this rapidly expanding field that we didn't touch upon. Below we collect a list of references for further reading. One could also consult the excellent reviews \cite{deLeeuwIpsenKristjansenWilhelm17, deLeeuw19} as well as the theses \cite{Widen19, Vardinghus19}. \\[6pt]
\indent An important open question concerns the extension of the results herein presented to strong coupling. A preliminary computation from string theory has been carried out in \cite{Buhl-MortensenLeeuwKristjansenZarembo15}, while one-loop corrections to the one-point functions of the supersymmetric D3-D5 system have been computed in \cite{Buhl-MortensenLeeuwIpsenKristjansenWilhelm16a, Buhl-MortensenLeeuwIpsenKristjansenWilhelm16c, Buhl-MortensenLeeuwIpsenKristjansenWilhelm17a} (see also the master's thesis \cite{Guo17}) and in \cite{GimenezGrauKristjansenVolkWilhelm19} for the non-supersymmetric D3-D7 system. Recently the paper \cite{GomborBajnok20} appeared together with an approach via supersymmetric localization \cite{Wang20, KomatsuWang20} (see also its precursors \cite{RobinsonUhlemann17, Robinson17}). Also quite recently the paper \cite{KristjansenMullerZarembo20} set up the computation of one-point functions in sections that contain gluon and fermion fields. See also \cite{OkamuraTakayamaYoshida05} for a study of the properties of spinning strings in the context of the AdS/{\color{red}d}CFT correspondence. \\[6pt]
\indent For applications to Wilson loops and lines see \cite{NagasakiTanidaYamaguchi11, deLeeuwIpsenKristjansenWilhelm16, AguileraDamiaCorreaGiraldoRivera16, PretiTrancanelliVescovi17, BonanseaDavoliGriguoloSeminara19, BonanseaIdiabKristjansenVolk20}.\footnote{Note also the relation of Wilson loop calculations to holographic entanglement entropy (cf.\ \cite{SeminaraSistiTonni18}).} The calculation of two-point functions and various applications to the conformal bootstrap program was taken up in \cite{deLeeuwIpsenKristjansenVardinghusWilhelm17, Widen17}. The non-supersymmetric non-integrable $SU(2)\times SU(2)$ symmetric dCFT that is dual to the D3-probe-D7 system was examined at tree and one-loop order in \cite{GimenezGrauKristjansenVolkWilhelm18, deLeeuwKristjansenVardinghus19}, while the beta-deformed D3-probe-D5 system is discussed in \cite{Widen18}. Many more holographic defect CFTs that are based on probe-brane systems and could possibly be studied by similar methods are known, see for example \cite{ConstableErdmengerGuralnikKirsch02a, AmmonErdmengerMeyerOBannonWrase09, ErdmengerMelbyThompsonNorthe20}. The S-dual of the D3-D5 system is the D3-NS5 brane system,\footnote{The author is thankful to C.\ Bachas for an interesting discussion about this system.} the integrability of which has been studied in \cite{Rapcak15}. An interesting future direction is the formulation of a defect version for the ABJM theory,\footnote{The author thanks S.\ Penati for drawing his attention to this subject.} see \cite{SakaiTerashima13} for a relevant preliminary investigation. \\[6pt]
\indent The relation of $SU(2)$ matrix product states to the N\'{e}el state was first explored in \cite{deLeeuwKristjansenZarembo15} and soon after in \cite{FodaZarembo15}. More recently, the 3-point structure constants of certain determinant operators and a non-protected operator of $\mathcal{N} = 4$ SYM were expressed in terms of determinant formulas similar to the ones we have been presenting in this talk \cite{JiangKomatsuVescovi19a, JiangKomatsuVescovi19b, JiangPozsgay20}. The structure constants are interpreted as $g$-functions or boundary entropies which are measures of the boundary degrees of freedom in a QFT. The $g$-function is obtained from the overlap of boundary states with the ground state of the system; see \cite{KostovSerbanVu18b} for a study relative to the purposes of the present work. \\[6pt]
\indent The subject of boundary and defect CFTs (holographic or not) is currently at the focus of intense research activity, see e.g.\ the recent review \cite{Andreietal18} and the events \cite{RoyalSociety2017, KingsCollegeLondon2019, PerimeterInstitute2019, Mainz2020}. In terms of physical applications that are relevant to the present approach, we single out the proposal of the D3-D7 system as a holographic model of graphene \cite{Rey08} and topological insulators \cite{KristjansenSemenoff16}. The connection with quantum quenches that was analyzed in \S\ref{Section:IntegrableDefects} has sparked a fruitful exchange of ideas between the two communities, further references on this subject include \cite{BertiniTartagliaCalabrese18, Piroli18, PerfettoPiroliGambassi19, ModakPiroliCalabrese19, RobinsondeKlerkCaux19}. \\[6pt]
\section[Acknowledgements]{Acknowledgements}
\noindent The author is grateful to George Zoupanos and the organizers of the 2019 Corfu Summer Institute for the invitation to participate to the \textit{Conference on Recent Developments in Strings and Gravity}. The author is thankful to M.\ Axenides, M.\ de Leeuw, E.\ Floratos, C.\ Kristjansen and K.\ Zarembo. Special thanks to M.\ Axenides for his feedback after reading various parts of the manuscript. The author would also like to thank H.\ B.\ Nielsen for illuminating discussions during the conference. \\[6pt]
\indent The author also thanks C.\ Papadopoulos and C.\ Markou for granting him access to the computing facilities of the Institute of Nuclear and Particle Physics at NCSR "Demokritos" and J.\ Ambj\o rn for the access to his computer system at the Niels Bohr Institute . \\[6pt]
\indent The author's research has received funding from the Hellenic Foundation for Research and Innovation (HFRI) and the General Secretariat for Research and Technology (GSRT), in the framework of the \textit{first post-doctoral researchers support}, under grant agreement No.\ 2595.
\appendix
\section{Norm of Bethe states \label{Appendix:BetheStateNorm}}
\noindent We define three functions $\varphi$ as the logarithm of the $SO(6)$ Bethe equations \eqref{BetheEquations1}--\eqref{BetheEquations3}:
\begin{IEEEeqnarray}{ll}
\varphi_{1,i} &= -i \log \Bigg[\left(\frac{u_i - \frac{i}{2}}{u_i + \frac{i}{2}}\right)^L \prod_{j\neq i}^{M} \frac{u_i - u_j + i}{u_i - u_j - i} \prod_{k=1}^{N_+} \frac{u_i - v_k - \frac{i}{2}}{u_i - v_k + \frac{i}{2}} \prod_{l=1}^{N_-} \frac{u_i - w_l - \frac{i}{2}}{u_i - w_l + \frac{i}{2}}\Bigg] \\[6pt]
\varphi_{2,i} &= -i \log \Bigg[\prod_{l \neq i}^{N_+} \frac{v_i - v_l + i}{v_i - v_l - i} \prod_{k=1}^{M} \frac{v_i - u_k - \frac{i}{2}}{v_i - u_k + \frac{i}{2}}\Bigg] \\[6pt]
\varphi_{3,i} &= -i \log \Bigg[\prod_{l\neq i}^{N_-} \frac{w_i - w_l + i}{w_i - w_l - i} \prod_{k=1}^{M} \frac{w_i - u_k - \frac{i}{2}}{w_i - u_k + \frac{i}{2}}\Bigg],
\end{IEEEeqnarray}
where $u_i \equiv u_{1,i}, \ v_j \equiv u_{2,j}, \ w_l \equiv u_{3,l}$ ($M = N_1, \ N_+ = N_2, \ N_- = N_3$) and consider the following Jacobian, that is known as the norm matrix:
\begin{IEEEeqnarray}{c}
G \equiv \partial_J \varphi_I = \frac{\partial\varphi_I}{\partial u_J}, \label{NormMatrix1}
\end{IEEEeqnarray}
with
\begin{IEEEeqnarray}{lll}
\varphi_I &\equiv \left\{\varphi_{1,i},\varphi_{2,j},\varphi_{3,l}\right\}, \quad & i = 1,\ldots, M, \ j = 1,\ldots,N_+, \ l = 1,\ldots,N_- \\[6pt]
u_J &\equiv \left\{u_i,v_j,w_l\right\}, \quad & I,J = 1,\ldots,M + N_+ + N_-.
\end{IEEEeqnarray}
Supposing that the Bethe roots $u_J$ are paired as in \eqref{BalancedBetheRoots1}--\eqref{BalancedBetheRoots2} it can be shown that the norm matrix \eqref{NormMatrix1} takes the form
\begin{IEEEeqnarray}{c}
G = \begin{bmatrix}

    {\color{blue}\begin{bmatrix}
    {\color{red}\begin{bmatrix} A_1 & A_2 \\ A_2 & A_1 \color{red}\end{bmatrix}} &
    \begin{matrix} B_1 & B_2 & D_1 \\ B_2 & B_1 & D_1 \end{matrix} \\
    \begin{matrix} B_1^t & B_2^t \\ B_2^t & B_1^t \\ D_1^t & D_1^t \end{matrix} &
    \begin{matrix} C_1 & C_2 & D_2 \\ C_2 & C_1 & D_2 \\ D_2^t & D_2^t & D_3 \end{matrix}
    \end{bmatrix}} &
    \begin{matrix} F_1 & F_2 & H_1 \\ F_2 & F_1 & H_1 \\ K_1 & K_2 & H_2 \\ K_2 & K_1 & H_2 \\ D_4^t & D_4^t & H_3 \end{matrix} \\

    \begin{matrix}
     \ \ \begin{matrix} F_1^t & F_2^t \\ F_2^t & F_1^t \\ H_1^t & H_1^t \end{matrix} & \ \
    \begin{matrix} K_1^t & K_2^t & D_4 \\ K_2^t & K_1^t & D_4 \\ H_2^t & H_2^t & H_3^t \end{matrix}
    \end{matrix} &
    \begin{matrix} L_1 & L_2 & H_4 \\ L_2 & L_1 & H_4 \\ H_4^t & H_4^t & H_5 \end{matrix}

    \end{bmatrix}, \label{NormMatrix2}
\end{IEEEeqnarray}
where the submatrices correspond to the norm matrices of the $SU(2)$ and $SU(3)$ subsectors, cf.\ \cite{deLeeuwKristjansenMori16}. It can be proven that the determinant of the norm matrix \eqref{NormMatrix1}, \eqref{NormMatrix2} factorizes as follows
\begin{IEEEeqnarray}{c}
\det G = \det G^+ \cdot \det G^-, \label{NormMatrixFactorization}
\end{IEEEeqnarray}
where $A_{\pm} \equiv A_1 \pm A_2$ (and so on for $B_{\pm}$, $C_{\pm}$, $F_{\pm}$, $K_{\pm}$, $L_{\pm}$) and
\begin{IEEEeqnarray}{c}
G^+ = \begin{bmatrix}
A_+ & B_+ & D_1 & F_+ & H_1 \\
B_+^t & C_+ & D_2 & K_+ & H_2 \\
2D_1^t & 2D_2^t & D_3 & 2D_4^t & H_3 \\
F_+^t & K_+^t & D_4 & L_+ & H_4 \\
2H_1^t & 2H_2^t & 2H_3^t & 2H_4^t & H_5 \\
\end{bmatrix} \quad \& \quad G^- = \begin{bmatrix}
A_- & B_- & F_- \\
B_-^t & C_- & K_- \\
F_-^t & K_-^t & L_-
\end{bmatrix}. \label{NormMatrices4}
\end{IEEEeqnarray}
\eqref{NormMatrices4} agrees with the corresponding formulas in the $SU(2)$ and $SU(3)$ sectors, cf.\ \cite{deLeeuwKristjansenMori16}. We have checked the equivalence of the definition \eqref{NormMatrices2}--\eqref{NormMatrices3} with \eqref{NormMatrices4} for a large number of states. According to the thesis \cite{Escobedo12} the norm of the $SO\left(6\right)$ Bethe states \eqref{NestedBetheEigenstates}, \eqref{NestedWavefunction1}--\eqref{NestedWavefunction2} is given by the formula:
\begin{IEEEeqnarray}{c}
\mathfrak{N}\left(L,M,N_+,N_-\right) = \left\langle\Psi|\Psi\right\rangle = \det G \cdot \prod_{i = 1}^{M}\left(u_i^2 + \frac{1}{4}\right),
\end{IEEEeqnarray}
which also has the factorization property as a consequence of \eqref{NormMatrixFactorization}.

\noindent
\bibliographystyle{JHEP}
\bibliography{D:/Documents/1.Research/2.Athenes/2.TeXFiles/Bibliography/JHEP_Bibliography,D:/Documents/1.Research/2.Athenes/2.TeXFiles/Bibliography/Math_Bibliography}

\end{document}